\documentclass[
reprint,
superscriptaddress,
floatfix,
amsmath,
amssymb,
aps,
notitlepage
]{revtex4-1}
\usepackage[hidelinks=true]{hyperref}
\usepackage{xcolor}
\usepackage{caption}
\usepackage{siunitx}
\usepackage{float}
\usepackage{subcaption}
\usepackage{amsmath, amssymb}
\usepackage{mathbbol}
\usepackage{color}
\usepackage{graphicx}
\usepackage{dcolumn}
\usepackage{bm}
\usepackage{hyperref}
\usepackage{xcolor}
\definecolor{greenish}{RGB}{135,200,18}
\definecolor{reddish}{RGB}{174,20,100}
\definecolor{blueish}{RGB}{30,100,198}
\hypersetup{
	colorlinks   = true, 
	urlcolor     = reddish, 
	linkcolor    = blueish, 
	citecolor   = greenish 
}


\def\min{\text{min}}

\begin{document}
\title{Optimal policies for mitigating pandemic costs}
\author{M. Serra$^{\dagger}$}
\author{S. al-Mosleh$^{\dagger}$}
\author{S. Ganga Prasath}
\author{V. Raju}
\author{S. Mantena}
\author{J. Chandra}
\author{S. Iams}
\affiliation{School of Engineering and Applied Sciences, Harvard University, Cambridge MA 02143.}
\author{L. Mahadevan }
\email{lmahadev@g.harvard.edu}
\email{\\ \noindent $\dagger$ {equal contribution}}
\affiliation{School of Engineering and Applied Sciences, Harvard University, Cambridge MA 02143.}
\affiliation{Department of Organismic and Evolutionary Biology, Harvard University, Cambridge, MA 02138, USA}
\affiliation{Department of Physics, Harvard University, Cambridge, MA 02138, USA}
\date{\today}

\begin{abstract}
Several non-pharmaceutical interventions have been proposed to control the spread of the COVID-19 pandemic. On the large scale, these empirical solutions, often associated with extended and complete lockdowns, attempt to minimize the costs associated with mortality, economic losses and social factors, while being subject to constraints such as finite hospital capacity. Here we pose the question of how to mitigate  pandemic costs subject to constraints by adopting the language of optimal control theory. This allows us to determine top-down policies for the nature and dynamics of social contact rates given an age-structured model for the dynamics of the disease. 
Depending on the relative weights allocated to life and socioeconomic losses, we see that the optimal strategies range from long-term social-distancing only for the most vulnerable, to partial lockdown to ensure not over-running hospitals, to alternating-shifts with significant reduction in life and/or socioeconomic losses. Crucially, commonly used strategies that involve long periods of broad lockdown are almost never optimal, as they are highly unstable to reopening {and entail high socioeconomic costs}. Using parameter estimates from data available for Germany and the USA, we quantify these policies and use sensitivity analysis in the relevant model parameters and initial conditions to determine the range of robustness of our policies. Finally we also discuss how bottom-up behavioral changes can also change the dynamics of the pandemic and show how this in tandem with  top-down control policies can mitigate pandemic costs even more effectively.  
\end{abstract}

\maketitle



\section{Introduction}
As of July 2020, the virus SARS-CoV-2 has infected more than thirteen million and been responsible for more than half a million deaths globally, devastating communities, economies and societies along the way \cite{galeotti2020uer}.  In the absence of therapies and vaccines to combat the COVID-19 disease, the primary approach to mitigate these losses has been to minimize the rate of spread of the infection - transmitted primarily via the respiratory tract - by controlling social interactions. At an extreme, this has led to the complete lock-down of entire societies, reducing social contacts to a minimum required for essential services. While this strategy reduces the infection rate dramatically \cite{grenfell-china}, it is unsustainable over longer terms owing to the considerable economic and social losses that it eventually entails - from loss of productivity to the collapse of vulnerable communities. This raises the natural question: how can one run a viable society limiting the   life, social and economic costs of the pandemic, while maintaining essential services and constrained by finite resources such as hospital capacity? 

\indent Mathematical models of the pandemic and its control by limiting social interactions and/or changing individual and collective behavior can help us understand the range of plausible scenarios and interventions \cite{projectingToPost}. Naturally, any model and the strategies that it suggests are only as good as the assumptions that it is based on and the data that feed into it.  {Here, we approach this question with the aim not as much as to be able to predict the course of the pandemic, but instead to use a set of minimal models grounded in data to provide qualitative scenarios for policies that mitigate the costs of the pandemic and sharpen the question of how to compare different policies.}

\indent The dynamics of epidemics has been the subject of mathematical study for more than a century since the pioneering work of Ross, Kermack and McKendrick \cite{ross, kermack1927contribution}. The theoretical framework for the evolution of epidemics takes the form of either deterministic or stochastic integro-differential equations for the rates at which a population of susceptible (S), infected (I) and recovered (R) individuals vary in space-time \cite{bailey1975mathematical,daleygani1999,anderson1992infectious,keeling2005networks,keeling2011modeling}; the simplest form of these models is the well known SIR model \cite{mathOfinfDisease}. Using this model and its variants, optimal strategies for containment of an epidemic in the form of vaccination and/or isolation while discounting future costs and allowing for stochastic effects \cite{morton1974optimal, wickwire1977mathematical}  have been studied for nearly fifty years. In the context of the current pandemic, this thread has been revived to determine a range of non-pharmaceutical interventions (NPIs) in different minimal scenarios \cite{opt4} inspired by optimal control theory \cite{OptContrSethi}.

\indent In order to help generate efficient NPIs, a model needs account for (\textit{\textbf{i}}) the differential vulnerability of populations as a function of age \cite{keeling2011modeling} that also accounts for their differences in social contact rates \cite{keeling2011modeling}, (\textit{\textbf{ii}}) the costs due to morbidity, mortality and healthcare {(life)} costs as well due to socioeconomic factors driven by distancing measures, (\textit{\textbf{iii}})  constraints due to finite resources e.g. hospital beds and intensive care units (ICU) capacity, (\textit{\textbf{iv}}) the possibility of batching strategies, where people participate economically in separate shifts and (\textit{\textbf{v}}) the behavioral dynamics of people driven by knowledge of infections.  Here, we extend a classical epidemiological model to account for these features and pose and solve an optimal control problem to generate  policies for mitigating pandemic costs as a function of the relative weights associated with health and socioeconomic costs. \\
\begin{figure}
	\centering
	\includegraphics[width=0.46\textwidth]{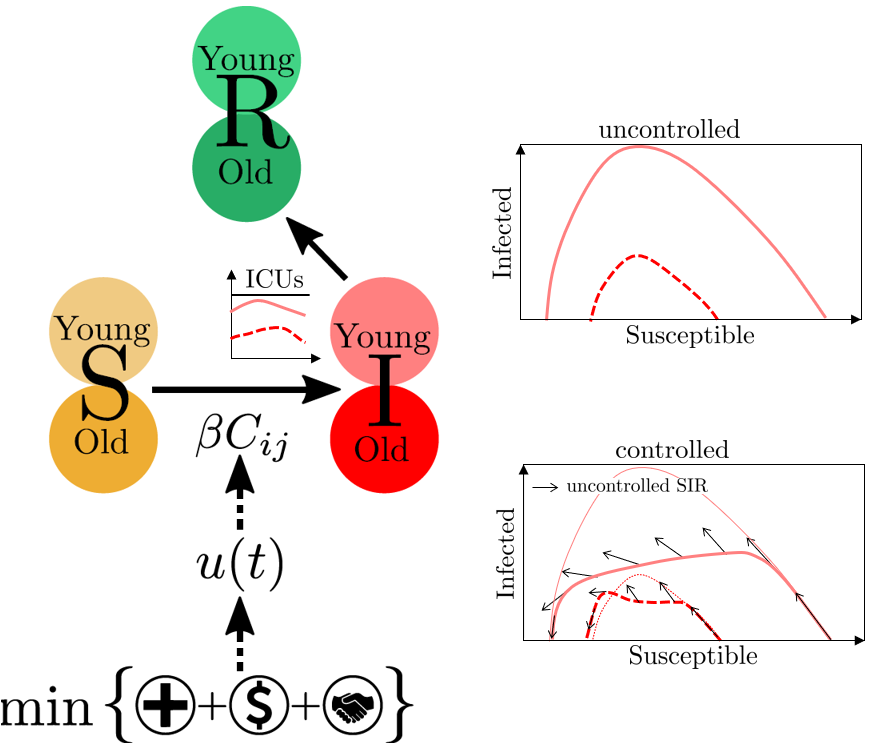}
	\caption{\textbf{Schematic of the 2-age SIR model.} $S_i$ is the susceptible population, $I_i$ the infected population and $R_i$ the recovered with $i=$ Young, Old. The controller $u(t)$ affects the dynamics of infected through $\beta C_{ij}$, where $C_{ij}$ is the contact matrix (Eq.~\ref{eq:ContactMatr}). The objective function we minimise is composed of life, economic and social costs (Eq.~\ref{eq:Jcost}). The right panels illustrate how $u(t)$ modulates the time evolution of infections in the young and old populations differently, optimally accounting for their intrinsic dynamics.}
	\label{fig:showing-compartments}
\end{figure}
\section{Mathematical framework}
\subsection*{Optimal control model}
Rather than using sophisticated spatio-temporal models that account for multiple compartments, stochasticity etc. but require multiple parameters (that still remain difficult to estimate from data), we modify the simple but effective SIR model to capture the essential features of the pandemic (See SI for an extension of our model to include a fourth compartment corresponding to the "exposed" individuals; our qualitative results are robust to this change). \\
\indent Our SIR model is assumed to have two epidemiological compartments $i=y,o$ ($y<$ 60 years; $o>$ 60) to account for the differential vulnerability, contact structure, infection and recovery rates in these sub-groups. Denoting by $S_i(t)$, the number of susceptible people in the age group labeled by $i$, $I_i(t)$, the corresponding number of infected, and $R_i(t)$, the number of recovered, the dynamical equations for their evolution is given by \cite{Keeling}:
\begin{align}
\dot{S_i} =& \ -\beta \sum_{j=y,o} S_i\;C_{ij}\; \frac{I_j}{N_j},
\nonumber \\ \dot{I_i} =& \ \beta \sum_{j=y, o} S_i\;C_{ij}\; \frac{I_j}{N_j}  - \gamma \;I_i, \label{eq:SIRMain}\\ 
\dot{R_i} =& \ \gamma \;I_i, \nonumber
\end{align}
where $C_{ij}$ represents the number of contacts a person of age $i$ makes with people from age $j$ (per day), $\beta$ corresponds to the nominal rate of infection from contacts and $\gamma$ is the nominal recovery rate, $N_i$ is the population of age group $i$ and we define the total population as $N = N_1 + N_2$, subject to the constraint $N_1C_{12} = N_2 C_{21}$ since the total number of contacts is symmetric.\\
\indent We assume that the main control measure available to policy makers is to enforce a reduction of the contact rate between individuals in different age groups. Then, if we take the contact matrix to be 
\begin{eqnarray}
C_{ij} = C^0_{ij} - u(t) \; C^C_{ij},  
\label{eq:ContactMatr}
\end{eqnarray}
where $C^0_{ij}$ represents the nominal contact matrix between people in the absence of control measures, and $C^C_{ij}$ encodes the relative change in contact structure imposed by the control function $u(t)$ that characterizes the magnitude of the lock-down. 
 {In our minimal framework, we assume that $u$ is a scalar time-dependent function, so that an age-structured social-distancing policy enters through the form of  $C^C_{ij}$, which we choose so as to reduce contacts with the older, more vulnerable population more strongly (See SI for details).} \\
\indent To determine the strategy $u(t)$, we need to define an objective cost function that must be minimized, and accounts for a life cost in terms of the proliferation of infections, a measure of economic cost (loss), and a social cost associated with the burdens due to social distancing measures. We further require that the total  number of people in critical condition (defined as a weighted fraction of those infected) is below the finite number of available hospital beds/ ICUs. 
Then, we may write the optimal control problem for the control $u(t)$ formally as follows: 
\begin{subequations}
	\begin{eqnarray}
	&\underset{u}{\mathrm{arg \;min}}&\; \int_0^T \overbrace{(G_{\text{life}} +G_{\text{econ.}} +G_{\text{social}}}^{G(\mathbf{x},u,t)})d t , \nonumber \\
	G_{\text{life}} &=& \alpha_L \; \left(\frac{p_{y} \; I_{y}(t) + p_{o} \; I_{o}(t)}{N_{ICU}}\right) \label{eq:Jcost} \\
	G_{\text{econ.}} &=& \alpha_E \; \left(1 - \frac{N - I_{y}(t) - I_{o}(t)}{N}(1 - u(t))\right) \nonumber\\ 
	G_{\text{social}} &=&  \alpha_S\;\left(\frac{u(t)}{u_M}\right)^2, \nonumber 
	\end{eqnarray}
	subject to the SIR model (\ref{eq:SIRMain}) and the constraints:
	\begin{eqnarray}
	 & I_C(t) \equiv \frac{p_y\;I_y(t)+p_o\;I_o(t)}{N_{ICU}} \leq 1, \notag\\
	 & 0 \leq u \leq u_M. \label{ineq:constraints}
	\end{eqnarray}
\end{subequations}
 Here, the first term $G_{\text{life}}$ is the cost associated with the expected fraction of people needing ICUs (relative to the total number of available ICUs ($N_{ICU}$)), where the parameters $p_y$ ($p_o$) are the (known) probabilities that an infected young (old) person will need an ICU. The second term $G_{\text{econ.}}$ is the economic cost associated with the loss in production capacity due to a reduction in the number of productive individuals. While it is possible to use more complex forms, e.g. the Cobb-Douglas function \cite{cobb1928theory}, this increases the number of parameters that we have to fit, and so we have chosen to use a simple linear form (See  SI for the results using a nonlinear Cobb-Douglas function with the same qualitative trends). The quantity $(1 - u(t))$ represents the fraction of people allowed to work \footnote{When the young and old populations are quarantined in different proportions, the expression for the fraction of people allowed to work would be slightly different (See SI). Since this detail does not change the nature of our results, the simpler expression given here suffices. } which is multiplied by the fraction of productive individuals (not infected). Finally, the social cost $G_{\text{social}}$ grows with social distancing and becomes larger with $u(t)$ relative to the maximum lock-down $u_M$ which defines the minimum residual contact rate between individuals possible, e.g. due to families (which can vary across cultures and societies \cite{projecting-contacts1}); we use a simple quadratic form to strongly penalize increase in $u(t)$ (See the SI for an exponential functional form which produces similar results to those presented here). The integrand to be minimized has three normalized costs with $\alpha_L$, $\alpha_E$ and $\alpha_S$ being the relative weights associated with life, economic and social costs. A complete derivation of the nonlinear differential equations associated with the optimal control problem obtained by the minimization of the constrained functional Eqs. (\ref{eq:Jcost}-\ref{ineq:constraints})  is given in the SI. There are a number of parameters in our problem, most of which can be estimated from data. From the perspective of the policy, there is freedom to vary the relative weights of life, economic and social costs $ \alpha_L, \alpha_E, \alpha_S,$ and the desired nature of the contact structure imposed by the lock-down $C_{ij}^C$.  Once these are chosen, the governing differential equations associated with optimal control (see SI) were solved using the Open Optimal Control Library (Open OCL) \cite{koenemann2017openocl}, which uses the nonlinear optimization tool CasADi \cite{CasADi}, via the MATLAB interface. 
\subsection*{Parameter estimation}
 From the perspective of optimal control theory, our aim is to estimate system parameters in absence of control measures, and then let the  policy modulate the effective  social contact structures within a given time horizon. This is to be contrasted with data-driven models that focus only on short-term predictions using model predictive control or its variants \cite{dehning2020inferring}. Our reasons for this are primarily associated with the present lack of fine-grained data that would allow for continuous adjustments in the controller. Instead, by using a classical optimal control framework and sensitivity analysis, our analysis provides longer-term robust policies. 
 \\
\indent The two nominal time scales in the problem are the infection and recovery rates $\beta, \gamma$ respectively, which we extract from publicly available data (see SI for details). In addition, we need to extract the  following quantities from data for solving the above control problem:
the nominal contact matrix $C^0_{ij}$, its leading eigenvalue $c^0$, the total number of people in each age-group $N_j$, the initial number of infected and recovered individuals $I_i(0), R_i(0)$, the probability of old and young people needing critical care $p_o,p_y$, the limits on the number of ICUs $(N_{ICU})$ and the maximum value of the control $u_M$. 
\indent Our parameter estimates focused on Germany because there are  publicly available age-structured datasets over a sufficiently long duration before and after the onset of lock-down \cite{german-data}. The data used for estimating the growth rate of the infected population is the time series of confirmed infected cases in Germany that captures the total of all currently active infections as well as recovered individuals, $I(t) + R(t)$. From this data set we extract the growth rate of the number of active infections in the early exponential growth phase. {\color{black} This quantity, given as $\rho = \beta \; c^0 - \gamma$,}
is related to the doubling time through $T_{doubling} = \log(2)/\rho$. For the German data, we find that the doubling time is about 3.5 days, i.e. $\rho = 0.2 \pm 0.03$. In order to estimate the basic reproduction number $\mathcal{R}_0=\beta\;c^0/\gamma$ \cite{R0-calculation},  we use its relation to the growth rate and the serial interval, the mean duration from the onset of symptoms of an infector to the onset in a person they infected \cite{SI-0}, to get $\mathcal{R}_0 \approx 2.2$ {(see SI for details)}. Using these values we find $\beta = 0.036$ and $\gamma = 0.16$. \\
\indent The German dataset also shows that the ratio between the young and old infected populations is approximately constant and is given by $I_y/I_o = 3.8$. This ratio reflects the dominant eigenvalue of the contact matrix and its left eigenvector (see SI for details). We use this ratio, along with the constraint $N_1 C_{12} = N_2 C_{21}$, to estimate the contact matrix given in the SI. When assessing the different control measures (Results Section), we will need the infection fatality rate (IFR) for the two age groups. Following the Centers for Disease Control and Prevention (CDC) data \cite{cdc-planning}, we estimate the IFRs to be $0.001$ and $0.02$ for the young and old populations. From \cite{cdc-planning}, we also find that $p_y = 0.0076$ and $p_o = 0.031$. These estimates were adjusted to account for an asymptomatic ratio of $35\%$ (see SI for details). Finally, ICU capacity estimates are taken from \cite{ICU-bed-data}. For Germany $N_{ICU} = 34 \times 10^{-5} N$ while for the U.S. $N_{ICU} = 26 \times 10^{-5} N$. To account for uncertainties we include a safety factor in these estimates and design our control policy using $ 0.8*N_{ICU} \geq p_y \; I_y + p_o\; I_o$.  By comparing the growth rate (of the infected population) in the phase before shelter in place to the phase afterwards in Germany, we find $u_M = 0.85$ \footnote{Since the the optimal control does not exceed $c = 0.6$ in all our solutions, the precise value of this upper bound will not affect our results.}. 
{When choosing $\mathbf{C^C}$ we assume that, when a fraction $u$ of the young population is in lockdown, a (bigger) fraction $u/u_M$ of the old population is (See SI).}
\section{Results} \label{sec:results}
 \subsection*{Top-down optimal policies}
{\indent We quantify the performance of the different policies associated with our choice of $ \alpha_L, \alpha_E, \alpha_S,$ with three different measures: $E_c = \frac{1}{\alpha_E \; T} \; \int_0^T \; G_{\text{econ.}}(t)\; dt,$  which quantifies economic loss and represents the fraction of days of lost economic activity, $N_D=  \frac{S_y(T) - S_y(0) }{N} \times 0.001 + \frac{S_o(T) - S_o(0) }{N} \times 0.02,$ which is the expected fraction of the population that will die after one year, computed from estimates of the infection fatality rate for the young ($0.001$) and the old ($0.02$) populations, and $T_C$, the time spent at peak hospital capacity. Table \ref{tab:CostsGermany2D} shows the performance of different strategies using these measures.}

\begin{figure}[!ht]
	\centering
	\includegraphics[width=0.8\columnwidth]{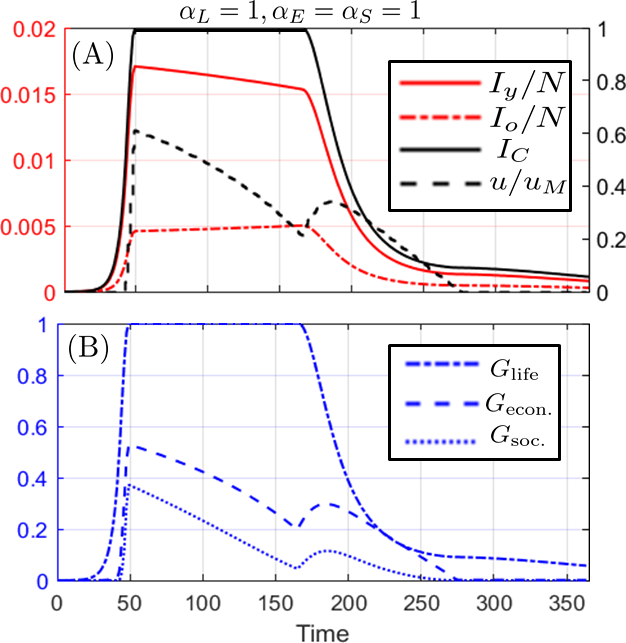}
	\caption{\textbf{Solutions of the optimal control problem}. (A) Solution to the optimal control problem described in \eqref{eq:Jcost} using the weights $\alpha_{L, E, S} = 1$. The quantity $I_C$ (\eqref{ineq:constraints}) represents the expected number of patients needing ICUs. The color of the plots determines the $y$-axis (left or right) they should be read from. (B) The values of the different costs ($G_{\text{life}}, G_{\text{econ.}}, G_{\text{soc.}}$) corresponding to panel (A). Parameters correspond to data from Germany (see SI). For analogous results using US data see the SI.} \label{fig:Germany2d}
\end{figure}

\indent As a benchmark, we first assumed equal weights, $\alpha_{L,E, S} = 1$ and calculated the optimal solution for this case. The results, shown in Figs.~\ref{fig:Germany2d}A-B suggest that the optimal controller starts with no lockdown i.e. $u = 0$, and increases with the rise in the number of infecteds,  reaching its maximum around day 50. After this initial growth, due to decrease in the number of susceptible people, the control measures gradually diminish over the course of about 175 days, but then increases again with a second peak soon after. This second peak is due to the fact that the number of older infected in the population slowly increases over the period of the decreasing partial lockdown, and eventually triggers an increase in the lockdown to prevent the number of older people who are infected from increasing further. This second peak in the lockdown makes the total period of saturated hospital capacity end sooner as can be observed. We emphasize that our scalar controller modulates the time evolution of infections in the young and old populations differently, optimally accounting for their intrinsic dynamics.  Figure \ref{fig:Germany2d}B shows the life, economic and social costs associated with this solution; the life cost tracks the weighted number of severely ill patients $I_C(t)$, the economic cost $G_{\rm econ.}$ tracks the control variable $u(t)$ since we have assumed a linear relationship linking them in \eqref{eq:Jcost}, and the social cost is quadratic in the control cost $u(t)$.  For the $\alpha_{L, E, S} = 1$ solution, we lose $18\%$ of economic activity with $0.3\%$ death rate after one year.
\begin{figure*}
	\centering
	\includegraphics[width=\textwidth]{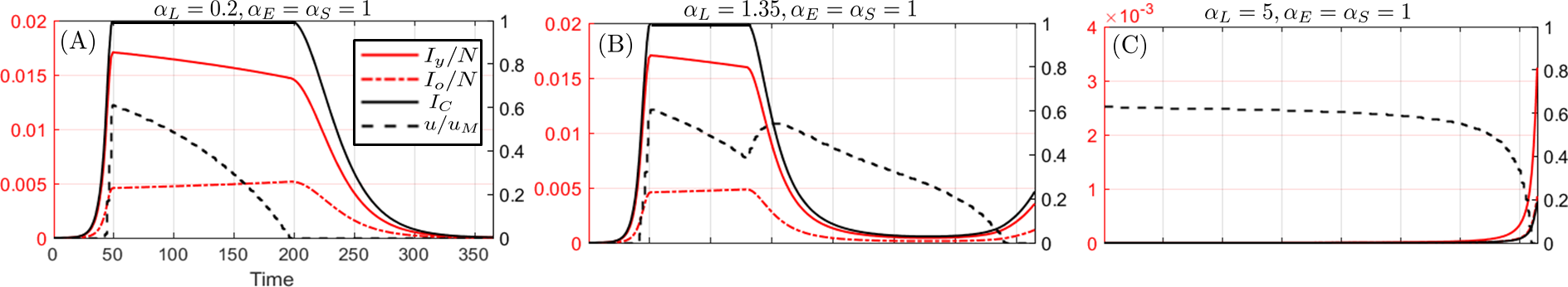}
	\caption{\textbf{Optimal control solutions for different weights on life and socioeconomic costs}. (A) Same as Fig. \ref{fig:Germany2d}A with high weights on the social and economic costs.  This strategy results in a shorter lockdown period combined with a longer period of maximum hospital capacity. (B) Increasing the weight on the life cost, $\alpha_L = 1.35$, leads to a longer lockdown period and a shorter period of maximum hospital capacity. (C) Continuing to increase the weight of the life cost leads to a constant lockdown (except near the end) to prevent the number of infections from increasing. The values of $\alpha_L$ are chosen to explore the different lockdown policies.} \label{fig:alpha-sweep}
\end{figure*}

\indent Moving away from the benchmark case of $\alpha_L=\alpha_E=\alpha_S=1$ and weighting the socioeconomic burdens relative to life costs changes the policies. In Fig.~\ref{fig:alpha-sweep}A we show that weighting the socioeconomic costs ($\alpha_L=0.2 < \alpha_E=\alpha_S=1$) strongly leads to the disappearance of the second peak in the social distancing control parameter $u(t)$, along with a corresponding reduction in duration of the control measures ({$E_c = 0.14$}) and a corresponding increase in the time spent at critical ICU capacity by $30\%$ (Table~\ref{tab:CostsGermany2D}). We note that increasing the socioeconomic weights much further does not change the qualitative nature of the solution significantly because of the resource constraint associated with ICU capacity. On the other hand, as we increase the relative weight on the life cost with ($\alpha_L=1.35 > \alpha_E=\alpha_S=1$), the resulting control policy $u(t)$ shown in Fig.~\ref{fig:alpha-sweep}B is similar to that shown in Fig. \ref{fig:Germany2d}A in the initial phase, starting at zero and then rising quickly. However, it will extend over a longer period of time and the second bump will be more pronounced, leading to a shorter time at maximum ICU capacity. Further increasing the weight to $\alpha_L = 5$ leads to a lockdown of nearly constant intensity (Fig.~\ref{fig:alpha-sweep}C). Note that considerably increasing $\alpha_L$ does not induce a much stronger lockdown. This is because it is sufficient to reduce the effective reproduction number to just below unity whence the intrinsic dynamics of disease transmission will limit the spread of the epidemic, and any farther increase in $u$ will just cause socioeconomic damages. A simple estimate of the maximum control required follows from the relation $(1 - u) \; \mathcal{R}_0 = 1$, leading to $u \approx 0.55$, which is close to the value observed in Fig.~\ref{fig:alpha-sweep}C. {This strategy results in low mortality ($N_D = 4 \times 10^{-5}$) and no strain on hospital capacity ($T_c = 0$), however, the economic burden will be great ($E_c \approx 0.5$). Therefore, we see that a strategy such as the $\alpha_{L,E,S} = 1$ solution can strike a balance between the two extremes in economic and life costs.} Finally, Fig. S1 shows the performance of a  periodic strategy with full lockdown ($\approx$ 2 months) followed by reopening ($\approx$ 2 months). This strategy leads to higher economic loss and considerably exceeds hospital capacity. 
\begin{table}[!htbp]
	\begin{tabular}{|c| c| c | c | c| c |}
		\hline
		& $\alpha_{L}=1$ & $\alpha_L= 5$ &  $ \alpha_{L}= 0.2$ & Periodic & Batching \\
		\hline
		\text{$E_c$} & 0.18  & 0.48 & 0.14 & 0.50 & 0.20 \\
		\hline
		\text{$N_D$} & $0.003$ & $4 \times 10^{-5}$ & $0.0036$ & - & $0.0017$  \\
		\hline
		\text{$T_c$} & 124 & 0 & 158 & - & 48 \\
		\hline
	\end{tabular}
	\caption{\label{tab:CostsGermany2D} \textbf{Comparing the performance of some of the control strategies shown in Figs.~(\ref{fig:Germany2d} - \ref{fig:compare-alternating})}. Here $E_c$ is the fraction of days of economic activity (per person) lost, $N_D$ is the expected mortality rate (fraction of dead in the population), $T_c$ is the time spent at peak hospital capacity. The periodic lockdown corresponds to SI, Fig. S1. The batching column corresponds to Fig.~(\ref{fig:compare-alternating}D) which may be compared with column $\alpha_{L} = 1$. Throughout our analysis we fix $\alpha_E = \alpha_S = 1$.} 
\end{table}
\subsection*{Contact allocation and batching} \label{sec:contact-allocation}
\begin{figure*}[t]
	\centering
	\includegraphics[width=\textwidth]{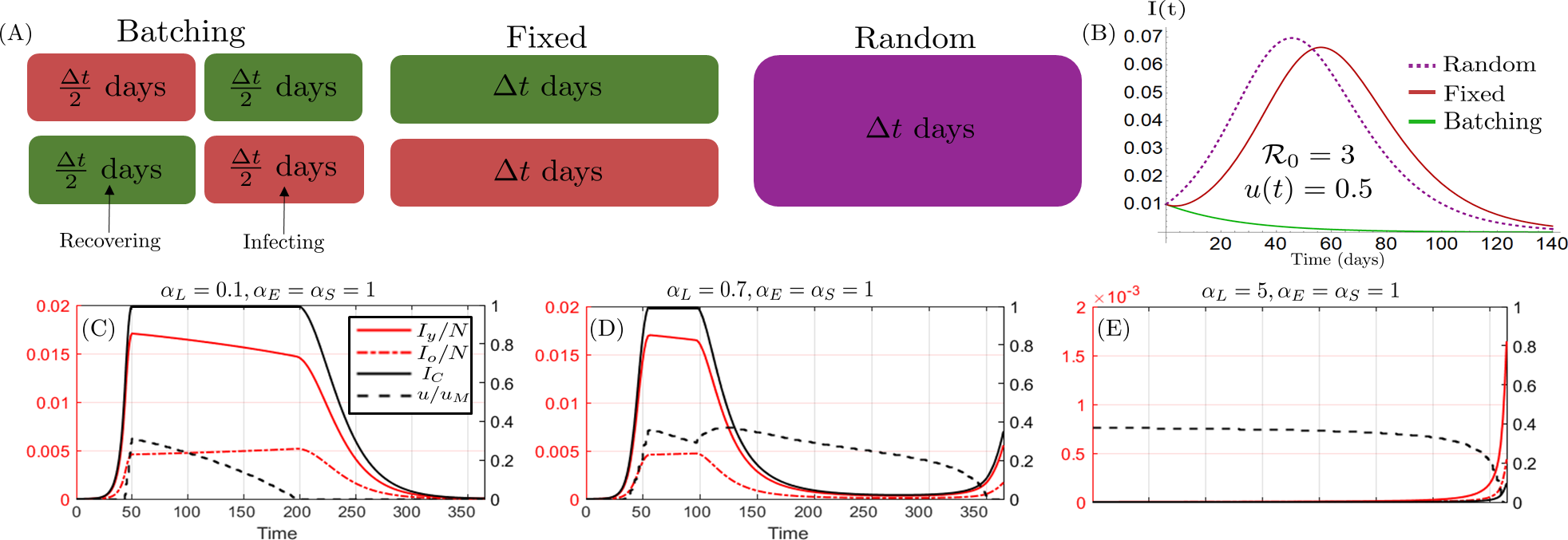}
	\caption{\textbf{Optimal batching strategies}. (A) The first row illustrates the reduction of transmission using batching strategies with participation ratio $(1 - u) = 0.5$. By letting each group recover for half of the shifting period, 3 days for example, and be active in the other, we reduce the effective reproduction number by a factor of 2. (B)  Comparison of the three different strategies for an epidemic with a basic reproductive number $\mathcal{R}_0 = 3$, $\gamma = 0.16$ and a fixed  $u(t) = 0.5$. Panels (C - E) show the solution of the optimal control problem with different values of $\alpha_L$ combined with the batching strategy, which effectively reduces $\beta$ by a factor of $(1 - u(t))$ as given in Eq.~(\ref{eq:effectGrowthRate}). Solutions obtained in this case achieve a better compromise between lives lost and economic loss. We can achieve lower economic impact with a slightly lower mortality (C) or save many lives without an exceedingly high economic cost (E). Compare panels C-E with Fig. \ref{fig:alpha-sweep}. {Similarly to Fig. \ref{fig:alpha-sweep}B, the values of $\alpha_L$ are chosen to explore the different lockdown policies. Parameters correspond to the German data set (see SI).}} \label{fig:compare-alternating}
\end{figure*}
Although the above framework provides the optimal value of the control variable $u(t)$, it doesn't specify how this can be realized in practice. The reduction in transmittance can be accomplished by reducing the number of contacts, or reducing the probability of infection per contact. The latter can be accomplished by masks, hygiene and other measures while the former can be accomplished by reducing the density of people in public and private gatherings. Naturally, reducing density can happen either through use of larger spaces or reducing the number of people in contact by a factor of $1- u$, which we denote as the participation number. Since it might not be feasible to enlarge the space of all gatherings, reduction in participation number is a necessary strategy to achieve a certain value of $u$. We now describe a solution to this allocation problem and illustrate how choosing the right strategies can result in a further reduction of infections and economic losses for the same $u$. \\
\indent For each strategy, we will take a given time period $\Delta t$, which for concreteness can be taken as one week, and divide it into two shifts. For example a given work week may be divided into two 3-day shifts. For simplicity, we also assume that at the beginning of each week the groups are selected from a homogeneous population (independently of age). More specifically, the ratio of susceptible people in each group at the beginning of the week is the same. This assumption, which may be relaxed, allows us to easily extend the previous optimal control problem for the case of batching strategies (see Fig.~\ref{fig:compare-alternating}). \\
\indent The first strategy involves having the same fraction of people ($N_1$) working for an extended period of time. At a given time $t$, the susceptible population is given by $S_1(t)$ and the infected is $I_1(t)$. We assume $S_1(t) \gg  I_1(t)$, which has to be true for a well controlled epidemic that is not near completion. Then, for a small enough period $\Delta t$ the number of infected increases by a factor
\begin{align}
I(t + \Delta t) =& \exp{\left[\gamma \left(\mathcal{R}(t) - 1\right) \Delta t\right]} \ I_1(t) + e^{- \gamma \Delta t} \ I_2, \nonumber \\
\mathcal{R}(t) \equiv &\ \frac{(1 - u(t)) \ S(t) \ \beta}{N(t) \ \gamma}. \label{eq:fixed}
\end{align}  
\indent In the second strategy, similar to that explored in \cite{CovidCyclicExit}, the two groups $N_1$ and $N_2$ alternate participation in periods of $\Delta t/2$ each. In time $\Delta t$, each group has participated a time $\Delta t/2$ and fully recovered with recovery rate $\gamma$ in a time period of $\Delta t/2$ (see Fig.~\ref{fig:compare-alternating}A). This combines to give the total number of infected as (see SI for details)
\begin{eqnarray}
I(t + \Delta t) = \exp{\left[\gamma \; \left(\frac{\mathcal{R}(t)}{2} - 1\right) \Delta t\right]} \;I(t). \label{eq:alternating}
\end{eqnarray}  
As an example,  suppose that at time $t = 0$ we split the population in half with $I_1 = I_2 = I(0)/2$. The ratio of increments in the number of infected will initially be given by
\begin{eqnarray}
\frac{2\ \exp{\left[\gamma \; \left(\frac{\mathcal{R}_0}{2} - 1\right) \Delta t\right]} \; }{\exp{\left[\gamma \; \left(\mathcal{R}_0 - 1\right) \Delta t\right]} + \exp\left[{- \gamma \Delta t}\right]} = \text{sech}\left(\frac{\gamma \; \mathcal{R}_0 \Delta t}{2} \right)<1. \;\;\;\;\;\; \nonumber
\end{eqnarray} 
 So for large $\mathcal{R}_0$ the alternating strategy represents a significant improvement over the constant fraction strategy (Fig.~\ref{fig:compare-alternating}B). \\
\indent For $u = 0.5$, we see that the alternating strategy effectively drops the reproduction number $\mathcal{R}_0$ by two (Eq.~\ref{eq:alternating}), while the constant strategy just amounts to a decrease in the initial number of actively infecting people by two (Eq.~\ref{eq:fixed}). Figure~\ref{fig:compare-alternating}B illustrates the difference between the two solutions over the course of an epidemic with $u = 0.5$ and compares them to the case of a completely mixed population. \\
\indent In the third strategy, corresponding to the completely mixed case there is no time for individuals to recover or discover symptoms while in confinement, and everyone is effectively always contributing to infection.
\begin{figure*}[t]
	\centering
	\includegraphics[width=1\textwidth]{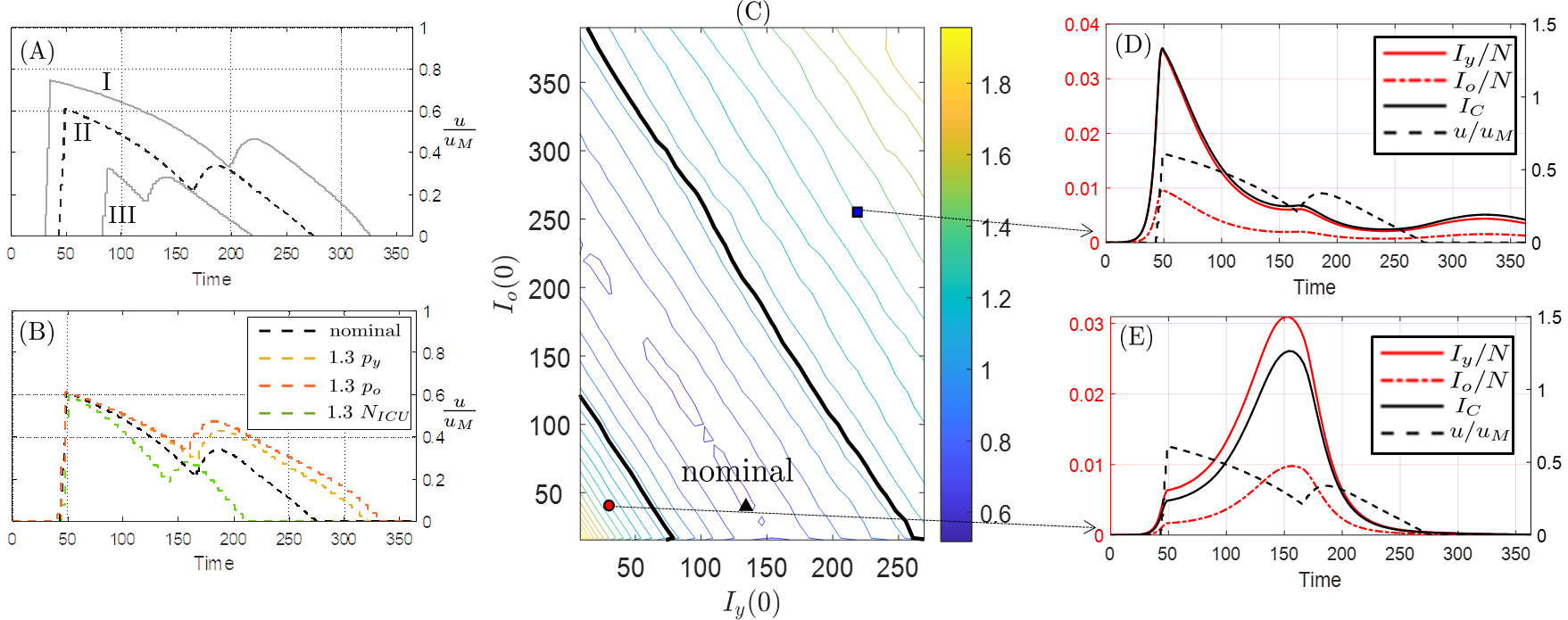}
	\caption{\textbf{Sensitivity of the optimal control strategy to parameters from German data estimates (see SI)} (A) Magnitude of the controller $u(t)$ subject to the worst and the best case estimates for the $\beta, \gamma$ parameter bounds. $I$ corresponds to the worst case ($\beta = 0.042,\ \gamma = 0.14$), $II$ corresponds to the nominal case ($\beta = 0.036,\ \gamma = 0.16$) shown in Fig.~\ref{fig:Germany2d}A, and $III$ to the best case ($\beta = 0.032,\ \gamma = 0.21$). (B) Variation of the controller to changes $p_y$, $p_o$ and $N_{ICU}$. 
	(C) Sensitivity of the controlled dynamics, using the nominal $u$, to uncertainties in the number of initial infected $I_y(0), I_o(0)$. Contours correspond to maximum value of $(p_y\ I_y(t)+p_o\ I_o(t))/N_{ICU}$, where $N_{ICU}=0.0003 \;N$. The uncertainty ranges of $I_y(0) (I_o(0))$ span large deviations from their nominal value marked by the black triangle. The black curves mark the critical level set beyond which the needed ICUs exceeds the available ones. Two of such cases are shown in panels (D-E).} \label{fig:Germany2dSens}
\end{figure*}
For arbitrary $u(t)$,  this strategy results in an effective growth rate (see SI)
\begin{eqnarray}
\lim_{{\Delta t} \rightarrow 0} \frac{I(t + {\Delta t}) - I(t)}{\Delta t \ I(t)} =  \gamma \ \left[ (1 - u(t)) \ \mathcal{R}(t)  - 1\right], \; \;\;\;\;\label{eq:effectGrowthRate}
\end{eqnarray}
which reduces the effective reproduction number (Eq.~\ref{eq:fixed}) by an extra factor of $(1 - u)$. \\
\indent With this result, we can implement the batching strategy in the optimal control framework described above. The only modification needed is to replace $\beta$ in Eq.~(\ref{eq:SIRMain}) with $(1 - u) \; \beta$ (SI for details). The results obtained are shown in Fig.~(\ref{fig:compare-alternating}). Note that the maximum lockdown required to stop the epidemic is now closer to $u = 0.33$ (Fig.~\ref{fig:compare-alternating}E) compared with $u = 0.55$ in the absence of batching (Fig.~\ref{fig:alpha-sweep}C). Furthermore, it is much easier in the present case to achieve a compromise between life and economic costs. Comparing the two socioeconomic focused strategies (Fig.~\ref{fig:alpha-sweep}A v.s. Fig.~\ref{fig:compare-alternating}C), we find that batching achieves a $40\%$ less economic losses. Comparing Fig.~\ref{fig:Germany2d}A with Fig. \ref{fig:compare-alternating}D we see a $50\%$ reduction in both the number of lives lost and time spent at maximum hospital capacity with a negligible increase in economic losses (Table \ref{tab:CostsGermany2D}). Our minimal approach on how to include the allocation problem into our optimal control framework already shows the resulting improvement in mitigating pandemic costs, at the slight expense of increasing the logistical complexity of batching. 

\subsection*{Sensitivity to parameters, cost functionals and epidemic model} \label{sec:sensitivity}
    Our results so far are driven by our choice of the epidemic model, the choice of cost functionals and are predicated on the parameters extracted from data. Understanding the range of robustness of our results to these choices requires us to vary each of these separately and determine their effect on the resulting policies.
    
    To understand the uncertainty in our parameter estimates, we perform a sensitivity analysis of the optimal control policy shown in Fig.~\ref{fig:Germany2d}A. The gray curves in Fig.~\ref{fig:Germany2dSens}A delimit the possible changes of the optimal control when $\beta, \gamma$ vary from the worst and the best case estimates obtained from data (see SI). This analysis shows that the shape of the nominal optimal control strategy (dashed black) is robust to uncertainties in $\beta, \gamma$. Figure \ref{fig:Germany2dSens}B shows the sensitivity analysis with respect to changes in the parameters $p_y,p_o,N_{ICU}$. A $30\%$ increase of $p_y$ induces a moderate increase in the optimal $u$, while a similar increase of $p_o$ leads to a larger change in lockdown intensity. By contrast, a $30\%$ increase of $N_{ICU}$ reduces the lockdown period by $\approx 70$ days as well as the overall lockdown strength. We note the robustness of the global shape of the optimal $u$ to changes in all parameters. \\
    \indent To quantify the sensitivity of the controlled dynamics to uncertain initial conditions, we consider a uniform grid of initial infected $I_y(0), I_o(0)$ spanning significant deviations from their nominal value marked by the black triangle. For each initial condition $I_y(0), I_o(0)$, we set $S_y(0) = N_y - I_y(0),\ S_o(0) = N_o - I_o(0), \ R_y(0)=R_o(0)=0$, and simulate the pandemic evolution using the nominal $u$. As a performance metric, for each initial condition we compute the maximum of $(p_y\ I_y(t)+p_o\ I_o(t))/N_{ICU}$, where $N_{ICU}=0.0003 \;N$, and plot the contour of this scalar field in Fig. \ref{fig:Germany2dSens}C. {\color{black} We note how these contours approximately run in the direction corresponding to $I_y(0) + I_o(0) = \text{constant}$, implying that our results are more sensitive to uncertainty in the total number of infected.} The black curves mark the critical (i.e. equal to 1) level set beyond which the needed ICUs exceed the available ones. Overall, our optimal controller guarantees that the number of available ICUs is enough for a large set of uncertainties in initial infected. In Figs. \ref{fig:Germany2dSens}D-E, we show the evolution of the pandemic in two cases where hospital capacity is exceeded.  Given a nominal control policy, underestimating the initial infected one expects a shortage of available ICSs (Fig.~\ref{fig:Germany2dSens}D). However, it is less intuitive that the same would happen when one designs the optimal $u$ overestimating the initial $I$ (Fig. \ref{fig:Germany2dSens}E). The reason behind this surprising result is that minimizing our costs tends to reduce $u$ to avoid unnecessary socioeconomic damages. Figure \ref{fig:Germany2dSens}E shows that starting from smaller $I_y(0), I_o(0)$, it takes longer to manifest a significant increase of infected, and by that time, the nominal decaying controller is unable to prevent exceeding hospital capacity. \\
 \indent To determine how our results change when using more complex models, we repeat our analysis by (i) replacing our SIR model by an SEIR model \cite{aron1984seasonality}, which incorporates an exposed but not yet infected group $E_i$, and (ii)  altering the socioeconomic cost functions as described (see SI for details). Neither of these chages the nature of our solutions (Figs.~S6 - S7).
 
\subsection*{Behavioral dynamics and bottom-up optimal policies}
\begin{figure*}[t]
	\centering
	\includegraphics[width=1\textwidth]{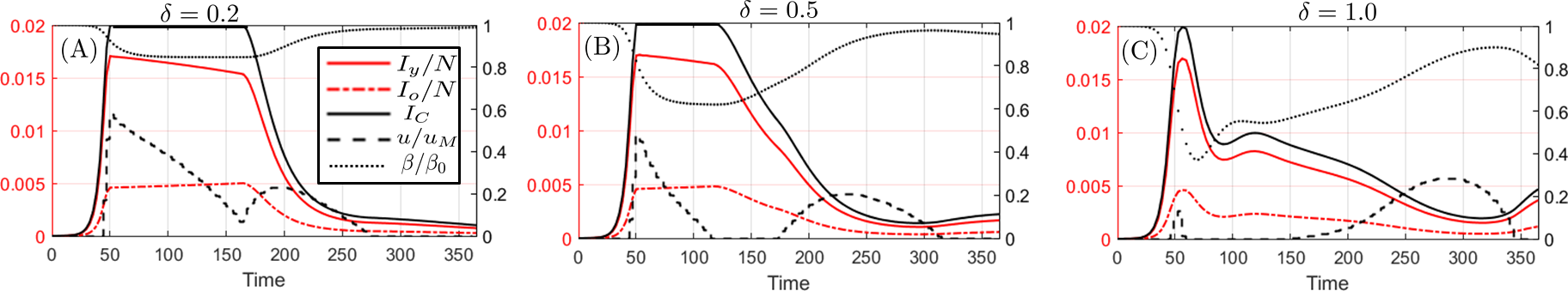}
	\caption{\textbf{Changing behavior in response to risk perception.} (A) Solution for the optimal control problem with behavioral dynamics incorporated (Eq.\ref{eq:behavioral-dynamics}). The parameters used are the ones extracted from German data (see SI), in addition to $\alpha_{L, E, S} = 1, \tau  = 2 \gamma^{-1}, v = 1$ and $\delta = 0.2$. (B \& C) Same as (A) with different values of $\delta$. Note how in (C), as the number of infections starts to decrease, the top-down control is required to increase to avoid further outbreaks.} \label{fig:behavioral-dynamics}
\end{figure*}

So far we have considered how the spread of infection may be curbed by externally imposed lockdown measures. However, the dynamics of disease transmission also critically depends on how people alter their behavior in response to perceived levels of risk \cite{adaptive-epidim}. To quantify this notion, we note the observation that as the number of (reported) cases goes up, without being forced to do so, people will often spontaneously practice more social distancing. This behavioral change can be minimally incorporated into the modified SIR model by making the overall contact rate parameter $\beta$ become a dynamical variable whose evolution follows the simple law
\begin{eqnarray}
\frac{d \beta(t)}{d t} = -\frac{\beta(t) -  \beta_0 \; \left(1 - \delta \;  \tanh\left[v \; I_C(t)\right]\right) }{\tau}, \label{eq:behavioral-dynamics}
\end{eqnarray}
where $\beta_0$ is the reference level at the start of the epidemic, the factor $\delta$ is a measure of the maximum change in $\beta$, $v$ determines the sensitivity of the behavioral response and $\tau$ is the time scale associated with the dynamics behavior change. While these parameters may be estimated by analyzing the effect of public events on mobility data \cite{grenfell-mobility}, this lies outside the scope of this paper.\\
\indent This extension of the SIR model allows us study the bottom-up response of the population to an evolving pandemic, and is particularly important for countries where the social costs of an enforced (top-down) lockdown can be high. Including \eqref{eq:behavioral-dynamics} in our control framework (see SI for details), we obtain different optimal scenarios by varying the parameter $\delta$, which represents the magnitude of the bottom-up behavioral response (Fig.~\ref{fig:behavioral-dynamics}). For small values of $\delta$ (Fig.~\ref{fig:behavioral-dynamics}A), the solution is similar to the results of Fig.~\ref{fig:alpha-sweep}, but with a slightly smaller enforced lockdown. As we increase $\delta$, the the optimal control $u(t)$ decreases in magnitude further until the two peaks become separated by a region having $u =  0$. This implies that when people respond strongly to a peak in the number of reported cases, there is no need for enforcing lockdowns from the top-down with the associated social costs. As $\delta$ increases further, the first peak goes away and we observe a later peak in the optimal policy (Fig. \ref{fig:behavioral-dynamics}C). This is because as the number of reported cases drops, $\beta$ increases, hence requiring a top-down intervention to prevent a second outbreak. 
\section{Discussion}
The value of a mathematical model is in its ability to (i) abstract a minimal framework that clearly lays out the underlying assumptions and (ii) use analysis combined with experimental data to provide qualitative insights that go beyond verbal reasoning. If these lead to a sharpening of the original question and direct further investigations, the model has served its purpose. We close with a discussion of the qualitative insights from the preceding calculations,  highlight the limitations of our model, and suggest possible future improvements to the question of how NPIs might minimize pandemic costs.
\subsubsection*{Age-structured partial lockdown outperforms periodic lockdown}
Taking into account the mortality and morbidity differences in addition to the difference in contact rates between and among the different age groups, we found optimal policies that better mitigate socioeconomic losses while reducing the life cost. {We emphasize that even using an easily implementable scalar controller these policies reduce contacts in the young and old age groups differently (Eq.\eqref{eq:ContactMatr}), accounting for their intrinsic dynamics (Figs. \ref{fig:Germany2d}-\ref{fig:compare-alternating})}.

\indent Solutions that heavily weight economic costs start with no lockdown ($u(t) = 0$) and only reduce contacts as the number of infections rises to near hospital capacity. Such a strategy results in a higher fraction of the population getting immunity, allowing the control to progressively drop down to zero over time. {This is to be contrasted with measures implemented by many countries, which start with a severe lockdown and then proceed to open up. Unless full lockdown is implemented for the entire duration of the pandemic (Fig.~\ref{fig:alpha-sweep}C), the risk of outbreaks remains extremely high when societies reopen (SI, Fig. S1), sadly being seen right now.} \\
\indent Complete lockdown ($u(t) = 1$) is not required in general to stop the spread of the epidemic, and instead it is enough to bring the effective reproduction number to a little less than unity. This is why in Fig. \ref{fig:alpha-sweep}C, with a higher weight on the life cost, the value of $u(t)$ does not exceed $ (\mathcal{R}_0 - 1) / \mathcal{R}_0 \approx 0.55$. 
\subsubsection*{Batching outperforms bulk strategies with marginally higher logistic costs} 
An effective SIR model that incorporates batching strategies into the optimal control framework lead to a $60\%$ reduction in the period of saturated hospital capacity and $50\%$ less mortality, with negligible increase of economic losses (Table \ref{tab:CostsGermany2D}). Furthermore, the minimum required value to suppress the epidemic in this case drops down to $u = (\sqrt{\mathcal{R}_0} - 1) / \sqrt{\mathcal{R}_0} \approx 0.33$ (see Fig.~\ref{fig:compare-alternating}E). In other words, about $70\%$ of people could be participating economically and working without any outbreaks, as long as $60\%$ of them are cycling in alternating shifts.
\subsubsection*{Bottom-up behavioral dynamics helps mitigating pandemic costs} Including bottom-up behavior changes due to risk perception in our optimal control problem shows that a significant reduction in hospital load can be achieved with mild imposed lockdowns (Fig. \ref{fig:behavioral-dynamics}C). This is due to the self-imposed isolation from people as the number of reported cases increases.

\subsubsection*{Sensitivity analysis quantifies robustness of policies to parameter uncertainty}  The nature of the optimal lockdown policy is robust to a range of uncertainty in the relevant model parameters and initial conditions (Fig.~\ref{fig:Germany2dSens}). Figure~\ref{fig:Germany2dSens}C illustrates the effects of wrongly estimating the initial infected on reaching hospital capacity. We find that both large underestimates  (Fig.~\ref{fig:Germany2dSens}D) and, interestingly, overestimates (Fig.~\ref{fig:Germany2dSens}E) of initial conditions can lead to a crisis driven by exceeding hospital capacity. The latter is because  the nominal controller is out of phase with the dynamics of the disease: it peaks too early and decays when the actual number of infected cases is increasing. \\
{\indent This last problem is exacerbated by delays in the onset of symptoms which makes using tests to estimate the current state of the system very difficult. Extrapolation, in addition to current test results, should be used to assess the current number of infections.} \\
\subsubsection*{Polcies are robust to form of socio-economic cost within a class of epidemic models}
{Our analysis is predicated on the classical SIR Model and a minimal cost that weights life, social and economic factors differentially. Replacing the SIR with the SEIR model, which incorporates an exposed but not yet infected group $E_i$ and using different forms of the socioeconomic cost does not change the qualitiative nature of our solutions, highlighting the robustness of our study (Figs.~S6 - S7).}
\subsubsection*{Accuracy of parameter estimation affects the robustness of optimal policies}
\indent While using real data allows to estimate the model parameters, and simulate and quantify the outcome of different optimal policies, the limited amount of data has made some parameters difficult to estimate, particularly those associated with infection rates. As more high quality data becomes available, our results may have to be updated in two ways: the nature of the optimal policies might change, and the range of robustness implied by sensitivity analysis will also vary.
\subsubsection*{Minimal framework does not account for stochasticity, additional compartments, spatial variability, complex batching and control}
Our simple SIR model with a two-age structure captures critical features in the dynamics of disease spread, such as the initial exponential growth, final herd immunity and their relation to the basic problem parameters. However, there are several effects that we have not included. The most important are to include the effects of additional compartments. Since COVID-19 is known to be transmitted through asymptomatic and presymptomatic individuals {(which may be a viral evolutionary adaptive strategy \cite{grenfell-asymptomatic})}, estimating the effect of these  could be an important addition to the present analysis. \\
\indent In addition, we have not explicitly accounted for stochasticity in disease transmission. Instead, we used a safety factor in $N_{ICU}$, and carried out a sensitivity analysis with respect to changes in the model parameters and initial conditions. In the early stages of the disease, given the importance of small number fluctuations, stochastic epidemic models \cite{Keeling, keeling2011modeling} might be more suitable.\\
 {Finally, adding more control inputs and complex batching of populations can lead to strategies that better exploit clinical and epidemiological differences between the different age groups and provide a better solution to the optimal allocation problem involving spatiotemporal interventions. Such changes, however, are likely to be more difficult to deploy.}\\
\section*{Acknowledgements}
We are grateful to Andrea Galeotti and Oliver Watson for illuminating discussions, and acknowledge support from the Schmidt Science Fellowship (MS) and the Postdoc Mobility Fellowship from the Swiss National Foundation (MS) and the Harvard QBIO Initiative (LM).
\clearpage

\onecolumngrid
	\clearpage
	\begin{center}
		\textbf{\large Supplemental Information}
	\end{center}
\appendix
  
\setcounter{figure}{0}
\setcounter{section}{0}
\setcounter{equation}{0}
\renewcommand{\thefigure}{S\arabic{figure}}
\renewcommand{\theequation}{S.\arabic{equation}}

\section*{Model}

Our model is a modification of the classical SIR model that accounts for an age-structured population with a non-trivial contact structure that follows the dynamics given by
\begin{align}
\dot{S_i} =& \ -\lambda_i(t) S_i, \label{eq:Sdot}\\ 
\dot{I_i} =& \ \lambda_i(t) S_i - \gamma I_i,\\
\dot{R_i} =& \ \gamma I_i, \label{eq:Rdot}\\
\lambda_i(t) =& \ \beta \sum_{j=1,2}C_{ij} \frac{I_j}{N_j}, \label{eq:lam}\\
C_{ij} =& \ C_{ij}^0 - u(t) C_{ij}^C\label{eq:C}.
\end{align}
To keep the model simple, we assumed only two age classes, and a scalar control parameter $u\in[0,u_M]$ that modulates the constant control-contact matrix $\mathbf{C^C}$ \eqref{eq:C}. For notation simplicity, we define the modified contact matrices $\mathbf{\tilde{C}}$ with entries $\tilde{C}_{ij} = C_{ij}/N_j.$ 
Using Eqs.(\ref{eq:lam}-\ref{eq:C}), we write $\pmb{\lambda}$ as 
\begin{equation}
\pmb{\lambda} = \beta\mathbf{\tilde{C}^0}\mathbf{I} - u\beta\mathbf{\tilde{C}^C}\mathbf{I},
\end{equation}
where $\mathbf{I} = [I_y,I_o]^\top$.
Denoting by $\mathbf{D^0}(\mathbf{x})$ the open-loop vector with entries $D^0_i=\beta \tilde{C}^0_{ij}I_j S_i$, and by $\mathbf{D^C}(\mathbf{x})$ the vector with entries $D^C_i=-\beta\tilde{C}^C_{ij}I_j S_i$, we can rewrite the dynamical system in compact form as
\begin{equation}
\label{eq:statespaceeqs_comp}
\underbrace{
	\left[\begin{array}{c}
	\mathbf{\dot{S}}\\
	\mathbf{\dot{I}}\\
	\end{array}\right]}_{\dot{\mathbf{x}}} = 
\underbrace{
	\left[\begin{array}{c}
	-\mathbf{D^0}(\mathbf{x})\\
	\mathbf{D^0}(\mathbf{x}) - \gamma\mathbf{I}\\
	\end{array}\right]}_{\mathbf{f^0}(\mathbf{x})} +u
\underbrace{
	\left[\begin{array}{c}
	-\mathbf{D^C}(\mathbf{x})\\
	\mathbf{D^C}(\mathbf{x})\\
	\end{array}\right]}_{\mathbf{f^C}(\mathbf{x})},\quad  \mathbf{f}(\mathbf{x},u) = \mathbf{f^0}(\mathbf{x}) + u \mathbf{f^C}(\mathbf{x}),\quad \mathbf{x}(0)=\mathbf{x_0}.
\end{equation}
Because $u_M<1$, if $\mathbf{C^0}=\mathbf{C^C}$, $C_{ij}>0$. To model a policy that favors contact inhibition of the old population using a scalar controller, we choose 
\begin{equation}
\label{eq:CcAgeDepPol}
\mathbf{C^C}= 
\left[\begin{array}{c c}
C^0_{yy} & C^0_{yo}/u_M\\
C^0_{oy} & C^0_{oo}/u_M
\end{array}\right]. 
\end{equation}
This corresponds to reducing the density of the young population by a factor of $(1 - u)$ and the old population by the smaller factor $(1 - u/u_M)$. For simplicity, we assumed that the rate of contacts with a population is proportional to the number of them not in lockdown, which leads to the given form of the $\mathbf{C^C}$. Note that when $u = u_M$, the second column of $C_{ij}$ will be zero, the minimum allowed value. In that case, the number of participating people in the old population has been reduced to zero. If the restriction in contacts is done differently, for example if both populations have the same participation ratio but older people are given stronger protective equipment, the expression for the matrix $\mathbf{C^C}$ will be different. \\
\indent The expression for the economic cost in the main text could be changed to account for the different rates of quarantine among the age groups. Specifically, 
\begin{eqnarray}
    G_{\text{econ.}} &=& \alpha_E \; \left(1 - \frac{N_y - I_{y}(t)}{N} \; (1 - u(t)) - \frac{N_o - I_{o}(t) }{N}\left(1 - \frac{u(t)}{u_M}\right)\right).
\end{eqnarray}
However, since this is not expected to change the qualitative nature of our results, we used the factor of (1 - u) for both populations in the economic cost. 
\section*{Optimal control} \label{sec:optContrApp}
In this section, we derive the set of equations to solve our optimal control {using the Pontryagin's Maximum Principle or indirect method} \cite{hartl1995survey,OptContrSethi}. {The derivation below applies to the optimal control problem described in Eq.~(3) of the main text and can be adjusted accordingly for the control problems involving batching or behavioral dynamics.}
\subsection*{Constraints and Lagrange multipliers}
In the language of optimal control, we have a Lagrange problem, with mixed inequality constraints
\begin{equation}
\mathbf{g}(u) = 
\left[\begin{array}{c}
u\\
u_M-u\\
\end{array}\right] \geq \mathbf{0},\ t\in[0,T], \label{eq:gconstr}
\end{equation}
and pure state inequality constraints
\begin{equation}
h(\mathbf{x}) = N_{ICU} - (p_y \; I_y(t)+p_o \;I_o(t)) \geq 0,\ t\in[0,T]. \label{eq:hconstr}
\end{equation}

Pure state constraints are usually more difficult to handle because they can be controlled only indirectly trough Eq~(\ref{eq:hRankCondDer}). We note that with a scalar controller it is typically not possible to enforce more than one pure state constraint as the corresponding full rank condition would not be satisfied.

The pure state inequality constraint is of order one, as $u$ appears for the first time in $h^1 = dh(\mathbf{x}(t))/dt = \langle \mathbf{\nabla_x}h,\mathbf{f}(\mathbf{x},u)\rangle$
\begin{equation}
\label{eq:hRankCondDer}
\mathbf{\nabla_{x}}h=\left[\begin{array}{cc}
0,0,-p_y,-p_o\
\end{array}\right];\quad 
h^1= \langle \mathbf{p}, -\mathbf{D^0}(\mathbf{x}) + \gamma \mathbf{I} - u \mathbf{D^C}(\mathbf{x}) \rangle \ ,
\end{equation}
where $\mathbf{p} = [p_y,p_o]^{\top}$ and $\langle ., .\rangle$ is the inner product between vectors.
With respect to the constraint $h(\mathbf{x})\geq 0$, an interval $(\theta_1, \theta_2) \subset [0, T]$ is called an interior interval if $h(\mathbf{x})>0,\ \forall \ t\in (\theta_1, \theta_2)$. If the optimal trajectory “hits the boundary,” i.e., satisfies
$h(\mathbf{x}, t)=0$, then $[\tau_1, \tau_2]$ is the
boundary interval. An instant $\tau_1$ is called an entry time if there is an
interior interval ending at $t=\tau_1$ and a boundary interval starting at
$\tau_1$. Correspondingly, $\tau_2$ is the exit time if a boundary interval ends
and an interior interval starts at $\tau_2$. If the trajectory just touches the
boundary at time $\tau$, while it is in the
interior just before and just after  $\tau$, then  $\tau$ is called a contact time. Taken
together, entry, exit, and contact times are called junction times. The pure state constraint is full rank on any boundary interval $[\tau_1,\tau_2]$ because 
\begin{equation}
\label{eq:hRankCond}
\text{rank}[\partial h^1/ \partial u] = \text{rank}\left[\begin{array}{c}
-\langle[p_y,p_o], \mathbf{D^C}(\mathbf{x}) \rangle
\end{array}\right]
=1
\end{equation}
from the definition of $\mathbf{D^C}(\mathbf{x})$. The mixed inequality constraint is also full rank because 
\begin{equation}
\label{eq:gRankCond}
\text{rank}[\partial \mathbf{g}/ \partial u, \text{diag} (\mathbf{g})] = 2
\end{equation}
along any optimal solutions. This full rank condition ensures that the gradients with respect to $u$ of all the mixed constraints are linearly independent.

The Lagrange multipliers must satisfy the complementary slackness condition
\begin{align}
&\mu_1\geq 0,\ \ \mu_1u = 0 \label{eq:slackcond_g1}\\
&\mu_2\geq 0,\ \ \mu_2(u_M-u)= 0 \label{eq:slackcond_g2}\\
&\eta\geq 0,\ \ \eta(N_{ICU} - (p_yI_y+p_oI_o))= 0, \ \dot{\eta}\leq 0. \label{eq:slackcond_eta2}
\end{align}

\subsection*{Solving the optimal control problem}

Using the indirect method maximum principle \cite{OptContrSethi}, we can then define the Hamiltonian and the associated Lagrangian as 
\begin{equation}
\label{eq:Ham}
H(\mathbf{x}, u, \pmb{\zeta}) =  \langle\mathbf{\pmb{\zeta}}, \mathbf{f^0}(\mathbf{x})\rangle +\langle\mathbf{\pmb{\zeta}}, u\mathbf{f^C}(\mathbf{x})\rangle -  G(\mathbf{x},u) ,
\end{equation}
\begin{equation}
\label{eq:Lag}
L(\mathbf{x}, u, \pmb{\zeta},\pmb{\mu}, \eta) =  \langle\mathbf{\pmb{\zeta}}, \mathbf{f^0}(\mathbf{x})\rangle +\langle\mathbf{\pmb{\zeta}}, u\mathbf{f^C}(\mathbf{x})\rangle -G(\mathbf{x},u) + \langle\mathbf{\pmb{\mu}}, \mathbf{g}(u)\rangle +\eta h^1(\mathbf{x}),
\end{equation}
where $\mathbf{\pmb{\zeta}}(t)$ is the adjoint vector, and $\mathbf{\pmb{\mu}}(t),\eta(t)$ the Lagrange multipliers associated to the inequality constraints. We note that maximizing $-G$ with respect to the control variable is equivalent to minimizing $G$ with respect to it.

From the maximizing condition $H(\mathbf{x^{*}}, u^{*},\pmb{\zeta}) \geq H(\mathbf{x^{*}}, u, \pmb{\zeta})$, the optimal controller $u^{*}$ 
\begin{equation}
u^*(\pmb{\zeta},\mathbf{x}^*)= \frac{u_M^2}{2 \alpha_S}\left[\langle \pmb{\zeta}, \mathbf{f^C}(\mathbf{x}^*) \rangle -\alpha_E\frac{N-I_y^*-I_o^*}{N}\right].
\label{eq:OptContr}
\end{equation}
In the interior of the feasible domain, i.e. where $h(\mathbf{x}^*)>0$, $\mathbf{g}(u^*)\geq 0$, while when $\mathbf{x^*}$ is on the boundary $\mathbf{x}^b$, $u^{b*}$ should satisfy the additional condition 
\begin{equation}
h^1(\mathbf{x}^b,u^{b*})\geq0,\quad \mathbf{x}^b=\{\mathbf{x}:\ h(\mathbf{x})=0 \}.
\end{equation} 

The differential equation for the adjoint vector $\mathbf{\pmb{\zeta}}$ is 
\begin{align}
&\dot{\mathbf{\pmb{\zeta}}}=-\frac{\partial L(\mathbf{x}^*, u^*(\pmb{\zeta},\mathbf{x}), \pmb{\zeta},\pmb{\mu}, \eta)}{\partial \mathbf{x} } = -[ \mathbf{\nabla}_{\mathbf{x}}\langle\mathbf{f}(\mathbf{x}^*,u^*(\pmb{\zeta},\mathbf{x}^*)),\mathbf{\pmb{\zeta}}\rangle]^{\top} +  [\mathbf{\nabla}_{\mathbf{x}}G(\mathbf{x}^*,u^*(\pmb{\zeta},\mathbf{x}^*))]^{\top} - [\mathbf{\nabla_{x}}h^1(\mathbf{x}^*,u^*(\pmb{\zeta},\mathbf{x}^*))]^{\top}\eta\label{eq:odeAdjoint}\\
&\mathbf{\pmb{\zeta}}(T^{-}) = [\mathbf{\nabla_{x}}h]^{\top}\gamma = \mathbf{0} \label{eq:odeAdjointTrsv1}
\end{align}
where $\gamma \geq 0,\ \ \gamma h(\mathbf{x^{*}},T)= 0$ and Eq.~(\ref{eq:odeAdjointTrsv1}) describe the transversality condition arising from the pure state constraint.

In the presence of inequality constraints, the optimal solution needs to satisfy additional conditions which will provide the remaining equations for the $\eta$ and $\pmb{\mu}$. To identify the ODE associated with the Lagrange multiplier $\eta$, we use the fact that along optimal trajectories $dH/dt=dL/dt=\partial L/\partial t$ which gives:
\begin{equation}
\frac{d}{dt}(\langle \pmb{\mu},\mathbf{g}(u^*))\rangle + \eta(t)h^1(\mathbf{x^*}))=0, \implies \langle \pmb{\mu},\mathbf{g}(u^*))\rangle + \eta h^1(\mathbf{x^*}) = \text{const.}.
\label{eq:etamueq}
\end{equation}

Additionally, the optimal trajectory $\mathbf{x}^{*}$ must also satisfy
\begin{align}
\frac{\partial L}{\partial u}\bigg\vert_{\mathbf{x}^{*},u^{*}(\pmb{\zeta},\mathbf{x}^{*})} &= \langle\mathbf{\pmb{\zeta}}, \mathbf{f^C}(\mathbf{x}^{*})\rangle - \partial_u G(\mathbf{x}^{*},u^{*}) + \langle\mathbf{\pmb{\mu}}, \partial_u \mathbf{g}(u^{*})\rangle  -\eta\langle\mathbf{p}, \mathbf{D^C}(\mathbf{x}) \rangle=0.\nonumber\\ 
&= \langle\mathbf{\pmb{\zeta}}, \mathbf{f^C}(\mathbf{x}^{*})\rangle - \partial_u G(\mathbf{x}^{*},u^{*}) + \mu_1-\mu_2 -\eta\langle\mathbf{p}, \mathbf{D^C}(\mathbf{x}) \rangle= 0\label{eq:dLduCond}.
\end{align}

From the complementary slackness conditions Eqs.~(\ref{eq:slackcond_g1}-\ref{eq:slackcond_eta2}) and Eq.~(\ref{eq:dLduCond}), the following equations hold along the optimal solution 
\begin{align}
& u^{*} = 0: \ \mu_2(t) = 0,\  \mu_1(t) = - \langle\mathbf{\pmb{\zeta}}, \mathbf{f^C}\rangle + \partial_u G  + \eta\langle\mathbf{p}, \mathbf{D^C} \rangle \label{eq:lagrMult1}\\
0<\  &u^{*}< u_{M}:\ \mu_1(t) = \mu_2(t) = 0,\ \eta\langle\mathbf{p}, \mathbf{D^C}\rangle = -\partial_u G + \langle\mathbf{\pmb{\zeta}}, \mathbf{f^C}\rangle\label{eq:lagrMult2}\\
& u^{*} = u_{M}:\ \mu_1(t) = 0,\  \mu_2(t) = \langle\mathbf{\pmb{\zeta}}, \mathbf{f^C}\rangle - \partial_u G - \eta\langle\mathbf{p}, \mathbf{D^C}\rangle \label{eq:lagrMult3}.
\end{align}
From \eqref{eq:slackcond_eta2}, when   $\mathbf{x}\notin \mathbf{x^b}$, $\eta(t)=0$ and the equations above fully determine $\pmb{\mu}$. When $\mathbf{x} \in \mathbf{x^b}$, Eqs.(\ref{eq:etamueq},\ref{eq:lagrMult1}-\ref{eq:lagrMult3}) determine $\pmb{\mu},\ \eta$. 
Therefore, Eqs. (\ref{eq:statespaceeqs_comp}, \ref{eq:OptContr}-\ref{eq:etamueq},\ref{eq:lagrMult1}-\ref{eq:lagrMult3}) completely define the boundary value problem that needs to be solved to compute $u^*$. Finally, at any junction time $\tau$, the following jump conditions need to be satisfied \cite{OptContrSethi}
\begin{align}
\mathbf{\pmb{\zeta}}(\tau^-)=& \ \mathbf{\pmb{\zeta}}(\tau^+)+ \alpha(\tau)[\mathbf{\nabla_{x}}h]^{\top} \label{eq:jumpCond1},\\
H(\mathbf{x^*}(\tau), u^{*}(\tau^-),\mathbf{\pmb{\zeta}}(\tau^-))=& \ H(\mathbf{x^*}(\tau), u^{*}(\tau^+),\mathbf{\pmb{\zeta}}(\tau^+)).
\end{align}
  Here we solve the optimal control problem numerically using the publicly available Open Optimal Control Library (Open OCL) \cite{koenemann2017openocl}, which effectively solves the optimal control problem using the direct method via Casadi \cite{CasADi}. 

\section*{Periodic strategies}
\begin{figure}[H]
	\centering
	\includegraphics[width=0.9\textwidth]{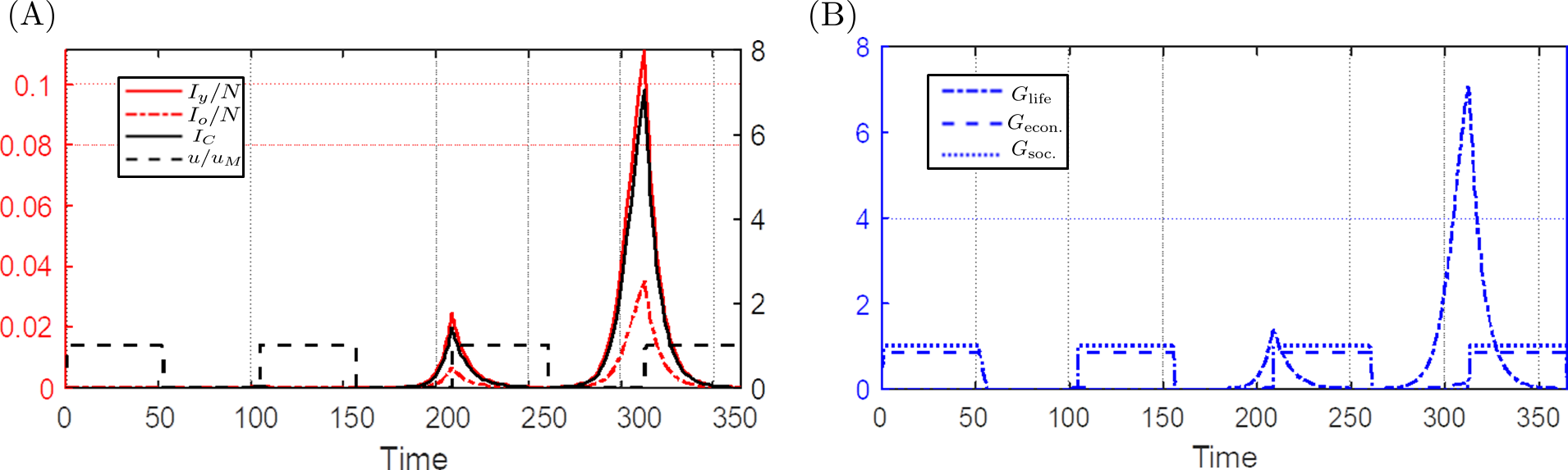}
	\caption{\textbf{Periodic Lockdowns.} Panels (A-B) Show the performance of a strategy that alternates between full lockdown ($u = 0.85$) and open ($u = 0$) states approximately every two months. While initially the lockdowns do suppress the outbreak, the situation is unstable and an outbreak goes out of control in the second open phase. This happens since the recovery rate during the lockdown is approximately $\gamma = 0.16$, while the growth rate for a population that is mostly susceptible (without control measures) is estimated to be $\rho = 0.2$. To prevent this resurgence either longer lockdown or milder re-openings will be required. The economic cost of this strategy, in terms of fraction of days lost will be greater $E_c \geq 0.5$. Since the maximum hospital capacity is exceeded for this solution, we do not calculate the period $T_c$ for this strategy.} \label{fig:shifts}
\end{figure}
\section*{Calculating the effective reproduction number for batching strategies}
We will consider here the case when the participation ratio (fraction of people participating economically, given by $ 1 - u$) is less that $0.5$. The case when $1 - u > 0.5$ can be solved in a similar fashion. We assume a strategy where a fraction (1 - $u$) of the population participates in the first time period ($\Delta t$ days) and another (1 - $u$) participates in the second time period, while a fraction $(2u - 1)$ does not participate in both periods. During the first time period, the number of infected grows as
\begin{eqnarray}
I \left(t + \frac{\Delta t}{2}\right) &=&  (2 \ u(t) - 1) \exp{ \left \{-\gamma \frac{\Delta t}{2} \right\} } \;I(t) + (1 - u(t)) \exp{ \left \{-\gamma \frac{\Delta t}{2} \right\} } \;I(t) \nonumber\\
&+& (1 - u(t)) \exp{ \left \{\gamma \; \left(\mathcal{R}(t) - 1\right) \frac{\Delta t}{2} \right\} } \;I(t).
\end{eqnarray}
In the second time period, the two groups switch places and we get
\begin{eqnarray}
I(t + {\Delta t}) =  (2 \ u(t) - 1) \exp{ \left \{-\gamma \Delta t \right\} } \;I(t)  + 2\ (1 - u(t)) \exp{ \left \{\gamma \; \left(\frac{\mathcal{R}(t)}{2} - 1\right) {\Delta t} \right\} } \;I(t).
\end{eqnarray}
We get the effective growth rate through 
\begin{eqnarray}
\rho_{eff} \equiv \lim_{{\Delta t} \rightarrow 0} \frac{I(t + {\Delta t}) - I(t)}{\Delta t \ I(t)} =  \gamma \ \left[ (1 - u(t)) \ \mathcal{R}(t)  - 1\right]. \; \label{eq:lambda-eff}
\end{eqnarray}
A similar calculation shows that for the case $u < 0.5$ we get the same expression for the effective growth rate. In that case a fraction of $(1  - 2 u)$ works in both periods. Two fractions of ratio $u$ work in alternative shifts (see Fig. \ref{fig:shifts}A). \\
\indent Fig.~\ref{fig:shifts}B shows a simulation of the effective description of the batching strategy compared with a more fine grained simulation that explicitly takes into account the shifts. Notice how the approximation becomes better for a smaller shift.  \\
\indent The derivations above assumed that $S$ changes slowly compared with $I$, which will be true if $S \approx N $. For example, significant variation in $I$ happens at a rate of order $\dot{I}/I \propto S/N$, whereas the rate of variation of S is much slower, $\dot{S}/S \propto I/N$. Consequently, when $\mathcal{R}_0 > 4$, the approximation above will not work for the entire duration of the simulation. 
\begin{figure}[H]
	\centering
	\includegraphics[width=0.9\textwidth]{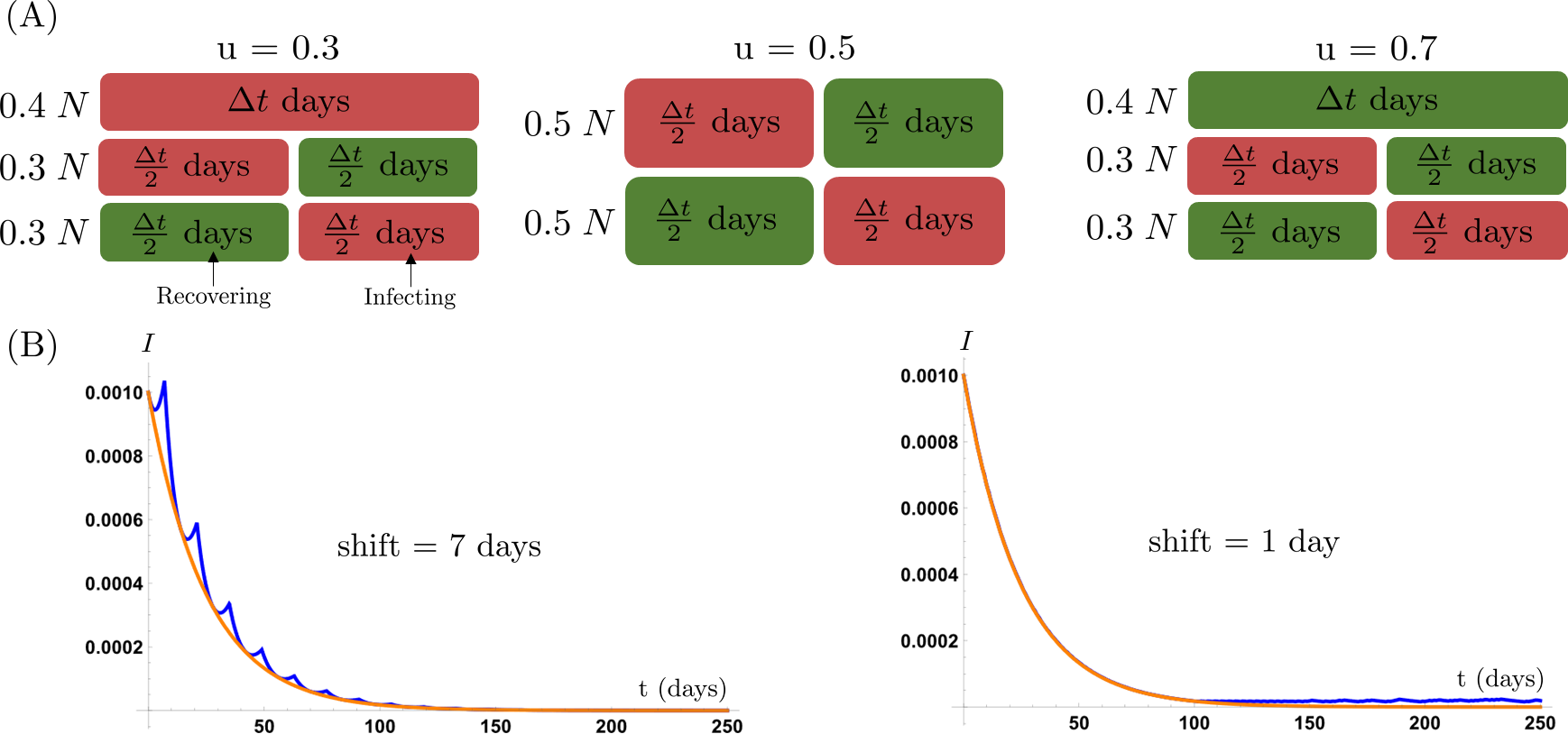}
	\caption{\textbf{Batching strategies} Panel (A) illustrates the batching strategies for three different values of $u$. For concreteness, we take $\Delta t = 7$ days. When $u < 0.5$, there will be a portion of the populations that is always infecting (left). By contrast, when $u > 0.5$, there will be a portion of the population that is always recovering. (B) Comparing the effective description of batching strategies to fine grained simulation. Here $\mathcal{R_0} = 3$ and $\gamma = 0.16$ and $u(t) = 0.5$. We assume initially $0.1\%$ of the population is infected and $R(0)  = 0$. } \label{fig:shifts}
\end{figure}

\section*{Parameter estimation for single and two population SIR model} \label{sec:parameter-est}
\subsection*{SIR without age structure}
\begin{figure}
	\centering
	\includegraphics[width=0.9\linewidth]{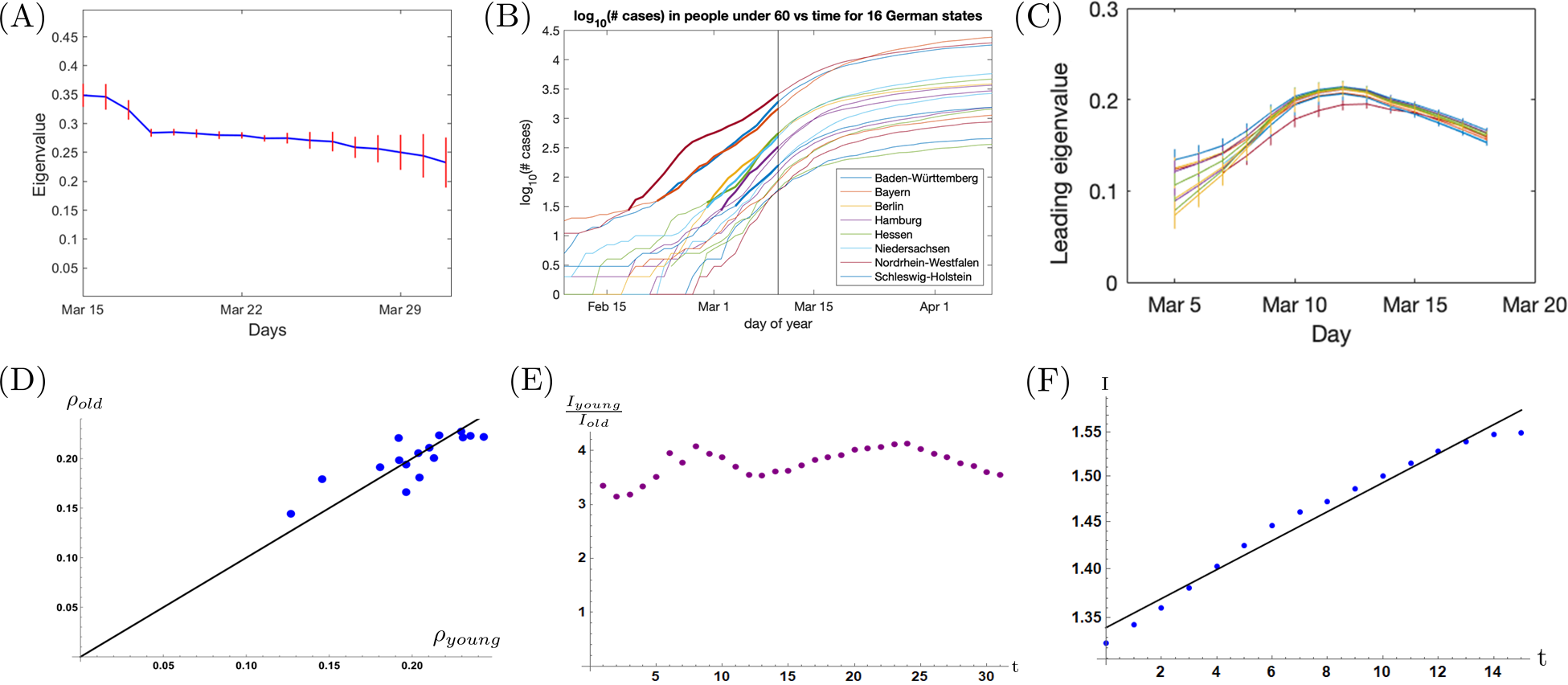}
	\caption{\textbf{Parameter estimation for U.S. and Germany}. (A) Eigenvalues (blue) determining growth of infections for U.S. national data is plotted with $95\%$ confidence intervals (red) against last day of a 15 day estimation period. Eigenvalues are shown for 17 such periods. (B) $\log_{10}$ of total cumulative cases in people ages 0 to 59 (left) is plotted against time for sixteen German states. The highlighted portions of the labeled curves were used for parameter estimation.  This data is selected from early in the virus spread, before the visible shift in growth rate that occurs around March 15th. The black vertical line marks the cutoff day of the data used for fitting. The starting day for each highlighted section is the day cases in that state crossed a threshold, set here to $40$ cases.  Parameter estimation was not sensitive to choosing a threshold of $100$ rather than $40$. (C)   This panel repeats the analysis of (A), which showed US data, for each German states separately. Note the interesting similarity of values between the different states. (D) As described in the text, the ratio of young to old should be dictated by the dominant eigenvalue of the contact matrix, in this figure we find a direct relation between the   growth rate of the infected in both young and old populations across the different German states (Blue points). This implies a constant fraction between the two age groups. (E) Shows the fraction of the populations in the different age groups as the number of infected increases by several orders of magnitude over the course of a month. (F) Log plot of the number of infected over time well after lockdown in Germany. The slope of the line is $\rho_1 = 0.2 \;\text{days}^{-1}$. } \label{fig:parameter-estimation}
\end{figure}

The data used for estimating the growth rate of the infected population is the total confirmed infected cases. This time-series captures the sum total of all currently active infected as well as recovered individuals, $I(t)+R(t)$. For the US \cite{US-data}, {we perform the estimation over a 15 day moving time window spanning the month of March (result in Fig.~\ref{fig:parameter-estimation}A). The maximum total detected cases was about $\approx 200,000$, less than $0.1\%$ of the total population and hence, the susceptible population is assumed to be a constant and equal to the total population ($S = N$) during the estimation period}. Under this assumption, we have linear dynamics for currently active infections and the recovered population:
\begin{equation}
\begin{array}{l}
\dot{I} = (\beta\; c  - \gamma) I, \;\;\;\;\;\;\;\; \dot{R} = \gamma \;I, \\
\end{array}
\end{equation}
where we factored out the mean number of contacts per day $c$ so that $\beta$ is the transmissibility. The solution is given as
\begin{equation}
\label{eqn:sirsoln}
\begin{array}{l}
I(t) = e^{(\beta\; c  - \gamma)\;t}I_0 \\
R(t) = R_0 + \gamma I_0(\beta\; c - \gamma)^{-1}\left(e^{(\beta\; c  -\gamma)\;t}-1\right)
\end{array}
\end{equation}
with $I(0) = I_0$ and $R(0) = R_0$.

To compare the trajectory generated by this model with the U.S. data for total infected cases, we sum $I(t)$ and $R(t)$ from Eq. \eqref{eqn:sirsoln} to get the expression
\begin{equation}
\label{eqn:sir}
I(t)+R(t) = a + b \;e^{\rho \; t}
\end{equation}
where $\rho = \beta\; c  - \gamma$, $a = R_0 - I_0/(\mathcal{R}_0 - 1)$ and $b = I_0\mathcal{R}_0 /({\mathcal{R}_0 - 1})$. Equation (\ref{eqn:sir}) captures the initial exponential growth phase in U.S. data. The uncertainty in initial conditions affects the estimated total cases through the constants $a, b$ whereas the eigenvalue $\rho$ captures the growth rate. Due to the dominant exponential growth term, uncertainty in the $I_0$ and $R_0$ via the constant $a$ has a comparably smaller effect on the fit. Given an estimate of $\rho$, we use an estimate of the serial interval \cite{SI-0, SI-1, SI-2} to get the complete parameter set. \\
\indent Thus if we fit $b$, $\rho$ and the initial total case count $H_0 = R_0 + I_0$ from data and obtain an estimate of the serial interval $\tau_s$, we can find the rest of the parameters as 
\begin{eqnarray}
I_0 = b \; \left(\mathcal{R}_0 - 1\right)/\mathcal{R}_0 , \;\;\;\;\; R_0 = H_0 - I_0, \;\;\;\;\; \nonumber\\
\beta\; c  = \frac{ \mathcal{R}_0 \rho}{ \mathcal{R}_0 - 1}, \;\;\;\;\; \gamma = \frac{\rho}{ \mathcal{R}_0 - 1}, \;\;\;\text{where} \;\;\;\; \mathcal{R}_0 \approx e^{\rho \; \tau_s}. \label{eq:params}
\end{eqnarray}
Fig.~(\ref{fig:parameter-estimation}A) shows the results for the fitting of U.S. data. The fit was done with the least square curve fit function in Matlab and the error bars are $95\%$ confidence intervals from this function. Since $\beta$ is always multiplied by $c$, when presenting the results for $\beta$ we assume $c = 10$ (Table~\ref{tab:germanySIR}).
\subsection*{SIR with age structure}
In this section, we describe the parameter estimation for the two age model based on the simplifying assumption that the ratio of infected young people to the infected old is constant throughout the period over which the parameters are estimated. This approximation works well since the leading (left) eigenvector of the contact matrix will dominate during this exponential growth phase. The next section will provide theoretical justification for this assumption while Figs.~(\ref{fig:parameter-estimation}C \& \ref{fig:parameter-estimation}D) show empirical justification in the case of Germany. Using the data \cite{german-data} illustrated in Fig.~(\ref{fig:parameter-estimation}D), we find that the mean ratio to be $I_{young}/I_{old} \approx 3.8$ and with standard deviation $0.27$. Furthermore, we find the magnitude of this leading eigenvector increases exponentially at the rate $\rho_1 = 0.2 \; \text{days}^{-1}$ (see Table \ref{tab:germanySIR}) which amounts to a doubling time of approximately $3.5$ days. Since we cannot estimate the contact matrix and the parameter $\beta$ independently, only their product, we normalize the contact matrix so that its leading eigenvalue is equal to 10, corresponding roughly to 10 contacts per day per person. \\
\indent Under an assumption of constant $S_y$, $S_o$, with $S_y\approx N_y$ and $S_o\approx N_o$, from Eqs.~(\ref{eq:Sdot} - \ref{eq:lam}), we have
\begin{equation}
\begin{array}{l}
\dot{I}_y = \beta\left(C_{yy} \;I_y + \dfrac{N_y}{N_o} C_{yo}\; I_o\right) - \gamma\;  I_y\\
\dot{I}_o =  \beta\left(\dfrac{N_o}{N_y} C_{oy}\; I_y  + C_{yy} \;I_o\right) -\gamma\; I_o\\
\dot{R}_y = \gamma\; I_y, \;\;\;\;\;\;\;\; \dot{R}_o = \gamma\; I_o \\
\end{array}
\end{equation}
We can rewrite these equations as
\begin{align}
\left[\begin{array}{c}
\dot{I}_y \\ 
\dot{I}_o \\ 
\dot{R}_y \\ 
\dot{R}_o 
\end{array}\right] &= \left[\begin{array}{cccc}
\beta \;{C}_{yy}-\gamma & \beta\; {C}_{oy} & 0 & 0 \\
\beta \;{C}_{yo} & \beta \;{C}_{oo}-\gamma & 0 & 0 \\
\gamma & 0 & 0 & 0 \\
0 & \gamma & 0 & 0 
\end{array}\right]     \left[\begin{array}{c}
{I}_y \\ 
{I}_o \\ 
{R}_y \\ 
{R}_o 
\end{array}\right] = M\left[\begin{array}{c}
{I}_y \\ 
{I}_o \\ 
{R}_y \\ 
{R}_o 
\end{array}\right].
\end{align}
The matrix $M$ has three distinct eigenvalues: $0$, $\rho_1$, $\rho_2$, with $\rho_1 > \rho_2 $, and $\rho_j$ has an associated eigenvector of the form $\underline v_j = \left[\begin{array}{c} 1 \ \  m_j \ \  \gamma/\rho_j \ \  \gamma \; m_j/\rho_j \end{array}\right]^T$, where $[1 \;m_j]$ are the left eigenvectors of the contact matrix whose eigenvalues $c_j$ satisfy $\beta \; c_j - \gamma = \rho_j$. The zero eigenvectors correspond to populations with only recovered people and zero infected. As shown in the next subsection, the linearized problem splits into two separate SIR models, one for each eigenvector of the contact matrix. \\
\indent We fit the data to the following equation, which only accounts for the leading eigenvalue of the contact matrix, 
\begin{equation}
\label{eqn:simplefitage}
\begin{array}{l}
I_o(t) + R_o(t) = a_y + b \;e^{\rho_1 t} \\
I_y(t) + R_y(t) = a_o + b \;m_1\;e^{\rho_1 t}
\end{array}
\end{equation}
As in the single age group case \eqref{eq:params}, we can use the exponential fit to estimate $b$ and $\rho_1$ in addition to $m_1$ and, using the initial populations numbers $I_i(0) + R_i(0)$, we can estimate the other parameters (Table \ref{tab:germanySIR}). Note that here $\mathcal{R}_0$ is estimated from the leading eigenvalue of the contact matrix $c^0$ through the relation $\mathcal{R}_0 = \beta \;c^0/\gamma$ (See \cite{R0-calculation}).
\begin{table}
	
	\begin{tabular}{|l| l| c | c | l |c | c | c | l |}
		\hline
		Country & $N_y$ & $N_o$ &  $\rho_1$ & $\gamma$ & $\beta$ & $\mathcal{R}_0$ & $\mathbf{C} = \left(\begin{array}{ c c} C_{yy} & C_{yo} \\ C_{oy} & C_{oo} \end{array}\right)$ \\
		\hline\hline
		Germany &  $57$ & $23$ & $0.2 \pm 0.03$& $0.16 \ (0.13 - 0.21)$ & $0.036 \ (0.032 - 0.042)$ & $2.2 \ (1.8 - 2.8)$ &$\left(\begin{array}{ c c} 8.5 \pm 0.4 & 2.3 \pm 0.2 \\ 5.6 \pm 0.6 & 1.5 \pm 0.4 \end{array}\right)$  \\
		\hline
		United States & $255$ & $74$ & $0.26 \pm 0.08$ & $0.14 \ (0.09 - 0.21)$ & $0.040 \ (0.032 - 0.048)$ & $2.8 \ (1.9 - 4.6)$ & $\left(\begin{array}{ c c} 7.4 \pm 0.3 & 2.3 \pm 0.2 \\ 8.2 \pm 0.9 & 2.6 \pm 0.7  \end{array}\right)$  \\
		\hline
	\end{tabular}
	\caption{\label{tab:germanySIR} Parameter estimates for the age-structured SIR model based on time-series data from Germany and the US. $\rho_1$ is estimated via curve fitting. $N_j$ is the populations in age group $j$, in millions of people. The parameters $\beta$, $\gamma$ and $\mathcal{R}_0$ are estimated using \eqref{eq:params}. The serial interval is taken to be $4 (3.5 - 4.5)$ days \cite{SI-1, SI-2}. The estimates of the growth rate are shown in Figs.~(\ref{fig:parameter-estimation}A  \& \ref{fig:parameter-estimation}B). Since we do not have time series age-structured data for the U.S. the error bars on the contact matrix are taken from the German counterpart.}
\end{table}

\subsection*{The dominant eigenvector of the contact matrix} \label{sec:contact}
In this section we show how to get an estimate of the effective contact matrix using its dominant left eigenvector. We write the age structured SIR model (for $S_j \approx N_j$) as
\begin{eqnarray}
\begin{pmatrix}
\dot{I}_{young} \\ \dot{I}_{old}
\end{pmatrix} \equiv \mathbf{\dot{I}} = \left(\beta\; {\mathbf{C}^T} - \gamma \right) \; \mathbf{{I}}, \;\;\;\;\;\;\;\;\;\;\;\;
\dot{\mathbf{R}} = \gamma \; \mathbf{I}. 
\end{eqnarray}
We can decompose both vectors $\mathbf{R}$ and $\mathbf{I}$ in terms of the eigenvectors of the contact matrix ${\mathbf{C}}$. Here, it is assumed that $\gamma$ is the same for all ages. If it's not then we can repeat the same analysis with the eigenvectors of the $2 \times 2$ matrix $\left(\beta\; {\mathbf{C}^T} - \boldsymbol{\gamma} \right)$, where $\boldsymbol{\gamma}$ is a diagonal matrix. \\
\indent Denoting the (left) eigenvectors of ${\mathbf{C}}$ as $\mathbf{V}_{\pm}$ and the eigenvalues as $\rho_{\pm}$, we find (because $\mathbf{V}_{\pm}$ are linearly independent) that the system of equations decouples for each eigenvector of the contact matrix. Specifically,
\begin{eqnarray}
\dot{I}_{\pm} = \left(\beta \; c_{\pm} - \gamma \right) \;I_{\pm}, \;\;\;\;\;\;\;\;\;\;\;\;\;\; \dot{R}_{\pm} = \gamma \; I_{\pm}.
\end{eqnarray}
where $I_{\pm}$ and $R_{\pm}$ are scalars and represent the components in the eigenbasis of the contact matrix. Each eigenvalue of the contact matrix determines two growth rates of the system. one of them is zero and the other is $\beta \;c_{\pm} - \gamma$. \\
\indent Thus, for long enough time, the system can be approximated by $\mathbf{I} \approx I_+ \; e^{(\beta\;c_{+} - \gamma) \; t} \;\mathbf{V}_{+}$. Thus the dominant eigenvector of the contact matrix should determine the long term fraction of infected people in the young and old populations. \\
\indent Conversely, if we have empirically that the ratio between the two populations is fixed over time (see Fig.~\ref{fig:parameter-estimation}C), we can use that to have a rough estimate of the contact matrix which is given by ${\mathbf{C}^T} = \rho_{+} \mathbf{V}_+ \mathbf{U}_+^T  +  \rho_{-} \mathbf{V}_- \mathbf{U}_-^T \approx  \rho_{+} \mathbf{V}_+ \mathbf{U}_+^T$, where $\mathbf{U}_\pm$ are the right eigenvectors of ${\mathbf{C}}$. Here we assume that $\mathbf{V}_{\pm}^T\cdot\mathbf{U}_{\pm} = 1$, otherwise we have to divide by the corresponding inner product in each term. \\
\indent From the German data, we have that $I_{young}/I_{old} = 3.8$ with standard deviation $\sigma = 0.3$. Thus we estimate the dominant eigenvector as $(3.8, 1)^T$ and estimating the left eigenvector through the consistency condition on the contact matrix we get the results shown in Tab. \ref{tab:germanySIR}. While there is no age structured time series data for the U.S. we estimate the dominant eigenvector using the aggregated data to be $(3.14, 1)$. 

\section*{Optimal lockdown policies for the US}
\begin{figure}[H]
	\centering
	\includegraphics[width=0.9\textwidth]{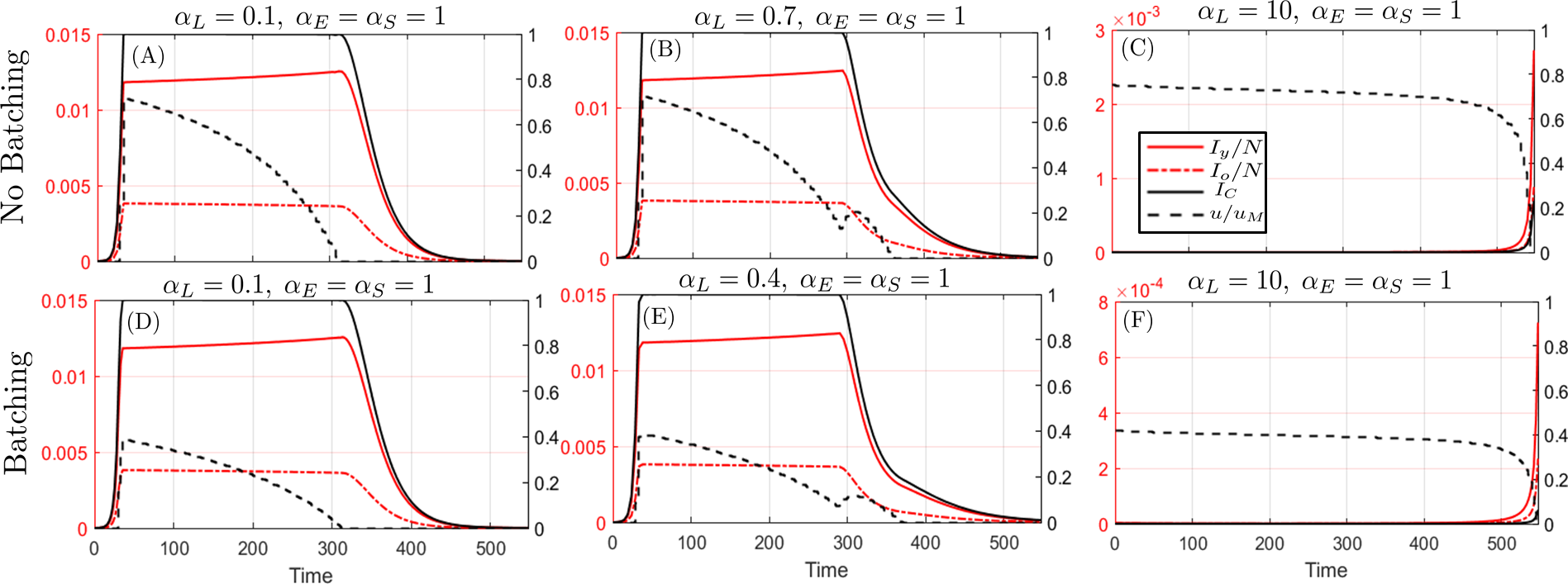}
	\caption{\textbf{Optimal control results for parameters extracted from U.S. data.} Panels (A - C) represent optimal control strategies with no batching (Similar to Fig. 3 in the main text). We have used a horizon period of 18 months. Due lower number of ICUs and a higher growth rate it is harder to reduce the period spent at hospital capacity and reach herd immunity within one year without extended lockdown measures. The situation is improved by incorporating batching strategies. Panels (D - F) show the behavior, for different values of $\alpha_L$, of the optimal batching strategies (Similar to Fig. 4 in the main text). } \label{fig:US2d}
\end{figure}

\section*{Behavioral dynamics}
\begin{figure}[H]
	\centering
	\includegraphics[width=1\textwidth]{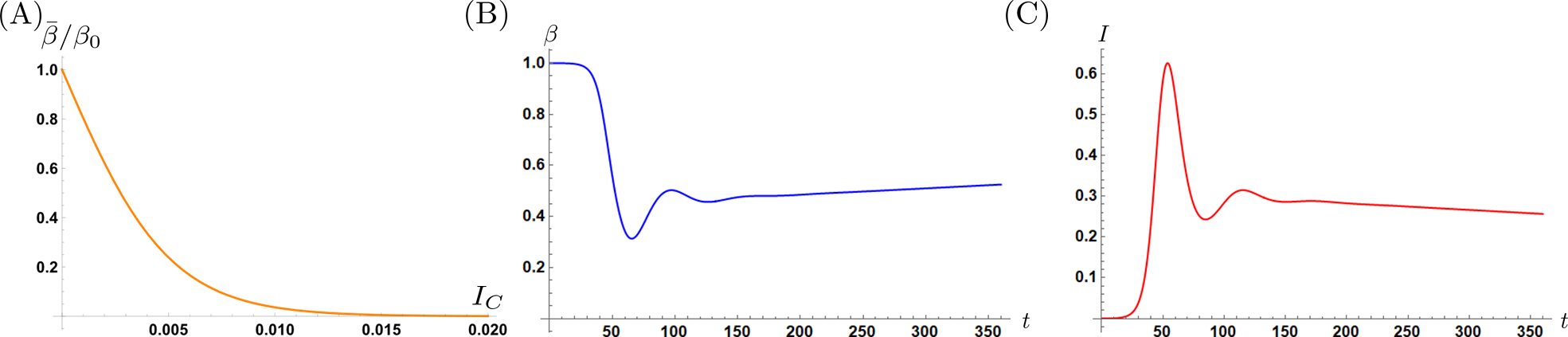}
	\caption{\textbf{Changing behavior in response to risk perception (open loop dynamics).} (A)  $\bar{\beta}$ is the value of $\beta$ that the population would settle on over time if $I_C$ was held fixed at a constant value, it is given by $\bar{\beta}/\beta_0 = (1 - \delta \;  \tanh\left(v \; I_C\right))$. The parameters used to generate this figure are $\beta = 0.036, \gamma = 0.16, c = 10, v = 2$ and $\delta  =1$. (B) Solution for $\beta$ as a function of time from \eqref{eq:beta-eq} and $u(t) = 0$. (C) The corresponding number of infections over time. Note how bottom up response of people can control the number of infections but will lead to oscillations.} \label{fig:behavioral-dyn}
\end{figure}

We incorporate the behavioral dynamics described in the main text by modifying the equations (S.1-S.5) to 

    \begin{eqnarray}
    \dot{S_i} &=& \ -\beta \sum_{j=y,o} S_i\;C_{ij}\; \frac{I_j}{N_j},
    \nonumber \\ \dot{I_i} &=& \ \beta \sum_{j=y, o} S_i\;C_{ij}\; \frac{I_j}{N_j}  - \gamma \;I_i, \nonumber \\ 
    \dot{R_i} &=& \ \gamma \;I_i, \nonumber\\
     C_{ij} &=& C^0_{ij} - u(t) \; C^C_{ij},  \\
    \dot{\beta} &=& -\frac{\beta(t) -  \beta_0 \; (1 - \delta \;  \tanh\left(v \; I_C\right)) }{\tau}. \label{eq:beta-eq}
    \end{eqnarray}
Note that here $\beta$ is a dynamical function rather than a constant. The solution to these equations for the open loop case, $u(t) = 0$, is given in Fig.~\ref{fig:behavioral-dyn}. \\ \indent  The objective function is changed (only the economic cost changes) to
	\begin{eqnarray}
	&\underset{u}{\mathrm{arg \;min}}&\; \int_0^T \overbrace{(G_{\text{life}} +G_{\text{econ.}} +G_{\text{soc.}}}^{G(\mathbf{x},u,t)})d t , \nonumber \\
	G_{\text{life}} &=& \alpha_L \; \left(\frac{p_{y} \; I_{y}(t) + p_{o} \; I_{o}(t)}{N_{ICU}}\right) \nonumber\\
	G_{\text{econ.}} &=& \alpha_E \; \left(1 - \frac{N - I_{y}(t) - I_{o}(t)}{N}(1 - u(t))\; \frac{\beta(t)}{\beta(0)}\right) \nonumber\\ 
	G_{\text{social}} &=&  \alpha_S\;\left(\frac{u(t)}{u_M}\right)^2, \nonumber 
	\end{eqnarray}
	subject to the constraints:
	\begin{eqnarray}
	 & I_C(t) \equiv \frac{p_y\;I_y(t)+p_o\;I_o(t)}{N_{ICU}} \leq 1, \nonumber\\
	 & 0 \leq u \leq u_M. \nonumber
	\end{eqnarray}

\section*{Model Sensitivity Analysis} 
In addition to studying the sensitivity of our predictions to uncertainties in the parameters, it would also be useful to know how our predictions would change for more realistic model. In order aid future work in this direction and to further test the robustness of the proposed strategies, we repeat our analysis by making two different changes to our basic model. 
\subsection{SEIR model} 
\begin{figure}[H]
	\centering
	\includegraphics[width=1\textwidth]{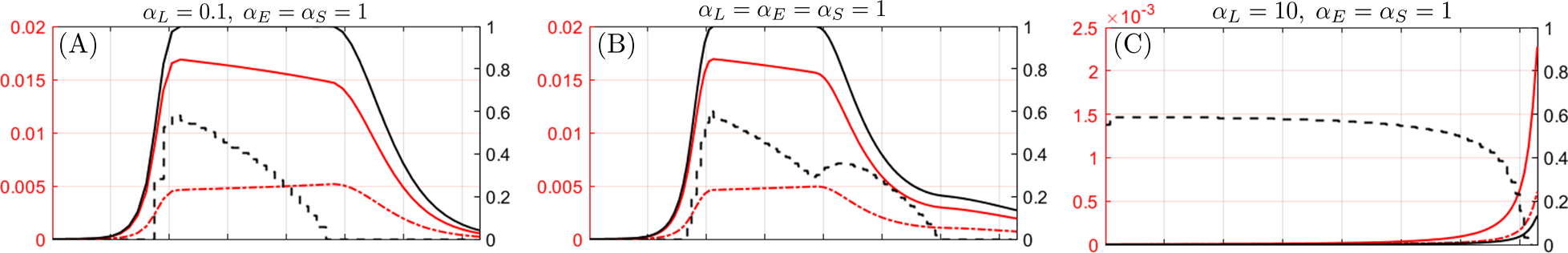}
	\caption{\textbf{Optimal controls for the SEIR model.} Solution of the optimal control problem with dynamics given in \eqref{eq:SEIR-model} and the cost function given by Eq.~(3) in the main text for (A) socioeconomically dominated (B) balanced and (C) life dominated cost functions. Compare with  Fig. 3 in the main text.} \label{fig:SEIR-analysis}
\end{figure}

The first change is to study an SEIR model \cite{aron1984seasonality} which incorporates an exposed but not yet infected group ($E_i$). The equations of motion in this case change to
\begin{eqnarray}
\dot{S_i} &=& \ -\beta \sum_{j=y,o} S_i\;C_{ij}\; \frac{I_j}{N_j},
\nonumber \\
\dot{E_i} &=& \ \beta \sum_{j=y, o} S_i\;C_{ij}\; \frac{I_j}{N_j}  - \sigma \;E_i,  \label{eq:SEIR-model} \\  
\dot{I_i} &=& \ \sigma \;E_i  - \gamma \;I_i, \nonumber\\ 
\dot{R_i} &=& \ \gamma \;I_i, \nonumber
\end{eqnarray}
where $\sigma^{-1}$ is proportional to the incubation period which we take to be four days so that $\sigma = 0.25 \;\text{days}^{-1}$. We repeat our optimization procedure with the same cost function and parameters used in Fig. 3 of the main text, and show our results in  Fig.~\ref{fig:SEIR-analysis}. We note that the results in this case are similar to those from the SIR model (Fig.~3 in the main text). 
\subsection{Alternative socioeconomic costs} 
\begin{figure}[H]
	\centering
	\includegraphics[width=1\textwidth]{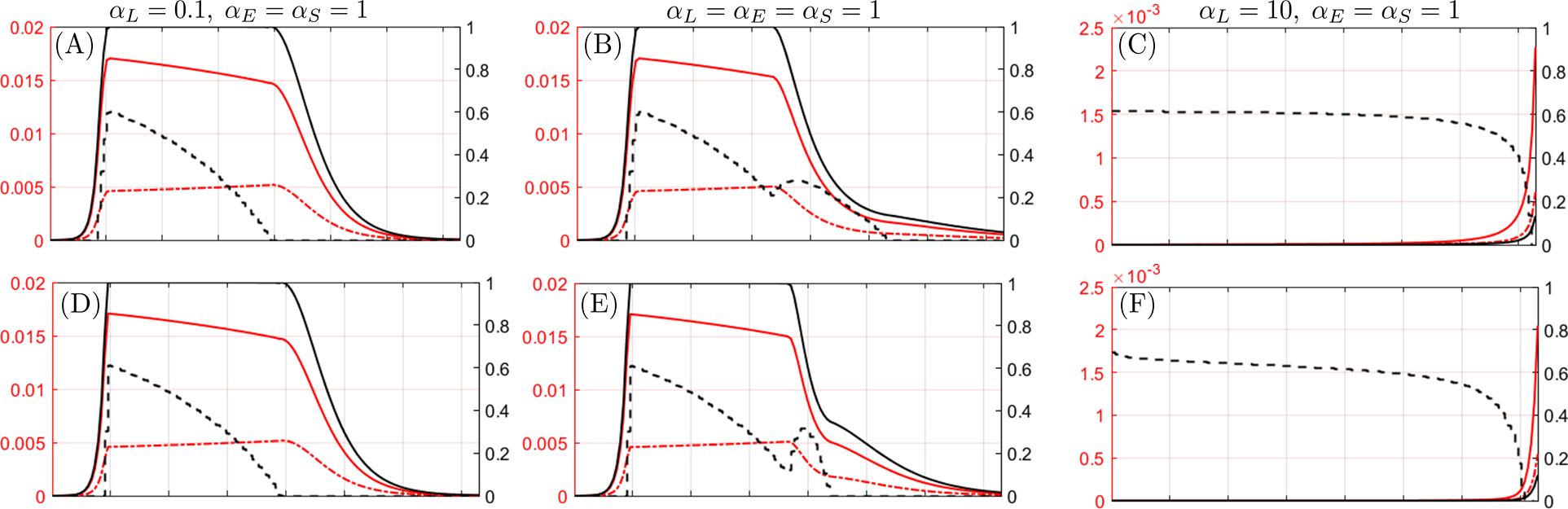}
	\caption{\textbf{Modifying the socioeconomic costs.} The first row represents the change of the quadratic dependence on $u$ of the social cost to $E_{\text{soc.}} \propto \left(e^{u/u_M}  - 1\right)^2$.
    The second row shows the result of changing the economic cost (using a Cobb-Douglas function with labor out put elasticity equal to 2). Compare with  Fig. 3 in the main text.} \label{fig:modified-costs}
\end{figure}

Here we analyse how our results change as we vary the choice of the socioeconomic cost function. We require the social cost to be increasing and convex with respect to $u$ so that it becomes steeper with higher lockdown measures. Fist, we modify the quadratic dependence on $u$ of the social cost to $E_{\text{soc.}} \propto \left(e^{u/u_M}  - 1\right)^2$. The results are shown in the first row of Fig.~\ref{fig:modified-costs} and again are consistent with the results of Fig.~3 in the main text. Second, we have modified the economic cost to the Cobb-Douglas function \cite{cobb1928theory} with labor output elasticity equal to 2. As shown in the second row of Fig.~\ref{fig:modified-costs}, the results are again similar to Fig.~3 in the main text. In this case, for $\alpha_L = 1$ the second bump has a much shorter duration due to the increased sensitivity of the economic cost on the fraction of people working. However, by increasing $\alpha_L$, thus emphasising the life cost more, the range of the second bump increases just as before. Therefore, even in this case, the range of policies observed remains qualitatively unaltered.


\section*{Glossary}
\begin{eqnarray*}
&\beta& \text{: \textit{Transmissivity}, Probability of an infection from a meeting of an infected and a susceptible person.}  \\
&\gamma& \text{: Rate of removal of infected individuals.}  \\
&C_{ij}& \text{: Number of contacts per day a person of age i makes with people from group j (contact matrix).}  \\
&c^{0}& \text{: Largest eigenvalue of the contact matrix.}  \\
&\mathcal{R}_0&  \text{: \textit{Basic reproduction number}, number of secondary cases one case would produce in a completely susceptible population. } \\ 
&I_i&  \text{: Number of infected people in group $i$. } \\ 
&S_i&  \text{: Number of susceptible people in group $i$. } \\ 
&R_i&  \text{: Number of Removed people in group $i$. } \\ 
&p_i&  \text{: probability of needing ICU for an infected person in group i. } \\ 
&N_{ICU}&  \text{: The number of available ICUs in the region of interest. } \\ 
&I_C&  \text{: The expected number of people needing ICUs as a fraction of total available ICUs. } \\ 
&u&  \text{: A function that controls the intensity of the lockdown measures. } \\ 
&\alpha_L&  \text{: The weight of the life cost in the objective function. } \\ 
&\alpha_E&  \text{: The weight of the economic cost in the objective function. } \\ 
&\alpha_S&  \text{: The weight of the social cost in the objective function. } \\ 
&\delta&  \text{: Magnitude of behavior change as response to change in infections. } \\ 
&v&  \text{: Sensitivity of behavior change to changes in the number of infections. } \\ 
\end{eqnarray*}






\bibliographystyle{ieeetr}

\begin{thebibliography}{10}

\bibitem{galeotti2020uer}
A.~Galeotti and P.~Surico, ``{A User Guide to COVID-19},'' {\em VOX CEPR Policy
  Portal}, vol.~27, 2020.

\bibitem{grenfell-china}
H.~Tian, Y.~Liu, Y.~Li, C.-H. Wu, B.~Chen, M.~U. Kraemer, B.~Li, J.~Cai, B.~Xu,
  Q.~Yang, {\em et~al.}, ``An investigation of transmission control measures
  during the first 50 days of the covid-19 epidemic in china,'' {\em Science},
  vol.~368, no.~6491, pp.~638--642, 2020.

\bibitem{projectingToPost}
S.~M. Kissler, C.~Tedijanto, E.~Goldstein, Y.~H. Grad, and M.~Lipsitch,
  ``Projecting the transmission dynamics of sars-cov-2 through the postpandemic
  period,'' {\em Science}, vol.~368, no.~6493, pp.~860--868, 2020.

\bibitem{ross}
R.~Ross, ``An application of the theory of probabilities to the study of a
  priori pathometry.—part i,'' {\em Proceedings of the Royal Society of
  London. Series A}, vol.~92, no.~638, pp.~204--230, 1916.

\bibitem{kermack1927contribution}
W.~O. Kermack and A.~G. McKendrick, ``A contribution to the mathematical theory
  of epidemics,'' {\em Proceedings of the Royal Society of London. Series A},
  vol.~115, no.~772, pp.~700--721, 1927.

\bibitem{bailey1975mathematical}
N.~T. Bailey, {\em The mathematical theory of infectious diseases and its
  applications}.
\newblock Charles Griffin \& Company Ltd, 5a Crendon Street, High Wycombe,
  Bucks HP13 6LE., 1975.

\bibitem{daleygani1999}
D.~J. Daley and J.~Gani, {\em Epidemic Modelling: An Introduction}.
\newblock Cambridge Studies in Mathematical Biology, Cambridge University
  Press, 1999.

\bibitem{anderson1992infectious}
R.~M. Anderson and R.~M. May, {\em Infectious diseases of humans: dynamics and
  control}.
\newblock Oxford university press, 1992.

\bibitem{keeling2005networks}
M.~J. Keeling and K.~T. Eames, ``Networks and epidemic models,'' {\em Journal
  of the Royal Society Interface}, vol.~2, no.~4, pp.~295--307, 2005.

\bibitem{keeling2011modeling}
M.~J. Keeling and P.~Rohani, {\em Modeling infectious diseases in humans and
  animals}.
\newblock Princeton University Press, 2011.

\bibitem{mathOfinfDisease}
H.~W. Hethcote, ``The mathematics of infectious diseases,'' {\em SIAM Review},
  vol.~42, no.~4, pp.~599--653, 2000.

\bibitem{morton1974optimal}
R.~Morton and K.~H. Wickwire, ``On the optimal control of a deterministic
  epidemic,'' {\em Advances in Applied Probability}, vol.~6, no.~4,
  pp.~622--635, 1974.

\bibitem{wickwire1977mathematical}
K.~Wickwire, ``Mathematical models for the control of pests and infectious
  diseases: a survey,'' {\em Theoretical population biology}, vol.~11, no.~2,
  pp.~182--238, 1977.

\bibitem{opt4}
R.~Chowdhury, K.~Heng, M.~S.~R. Shawon, G.~Goh, D.~Okonofua, C.~Ochoa-Rosales,
  V.~Gonzalez-Jaramillo, A.~Bhuiya, D.~Reidpath, S.~Prathapan, {\em et~al.},
  ``Dynamic interventions to control covid-19 pandemic: a multivariate
  prediction modelling study comparing 16 worldwide countries,'' {\em European
  Journal of Epidemiology}, pp.~1--11, 2020.

\bibitem{OptContrSethi}
S.~P. Sethi, {\em {Optimal Control Theory}}.
\newblock No.~3, Springer International Publishing, 2019.

\bibitem{Keeling}
C.~E. Dangerfield, J.~V. Ross, and M.~J. Keeling, ``Integrating stochasticity
  and network structure into an epidemic model,'' {\em Journal of The Royal
  Society Interface}, vol.~6, no.~38, pp.~761--774, 2009.

\bibitem{cobb1928theory}
C.~W. Cobb and P.~H. Douglas, ``A theory of production,'' {\em The American
  Economic Review}, vol.~18, no.~1, pp.~139--165, 1928.

\bibitem{Note1}
When the young and old populations are quarantined in different proportions,
  the expression for the fraction of people allowed to work would be slightly
  different (See SI). Since this detail does not change the nature of our
  results, the simpler expression given here suffices.

\bibitem{projecting-contacts1}
K.~Prem, A.~R. Cook, and M.~Jit, ``Projecting social contact matrices in 152
  countries using contact surveys and demographic data,'' {\em PLOS
  Computational Biology}, vol.~13, pp.~1--21, 09 2017.

\bibitem{koenemann2017openocl}
J.~Koenemann, G.~Licitra, M.~Alp, and M.~Diehl, ``Openocl--open optimal control
  library,'' 2017.

\bibitem{CasADi}
J.~A.~E. Andersson, J.~Gillis, G.~Horn, J.~B. Rawlings, and M.~Diehl,
  ``{CasADi} -- {A} software framework for nonlinear optimization and optimal
  control,'' {\em Mathematical Programming Computation}, vol.~11, no.~1,
  pp.~1--36, 2019.

\bibitem{dehning2020inferring}
J.~Dehning, J.~Zierenberg, F.~P. Spitzner, M.~Wibral, J.~P. Neto, M.~Wilczek,
  and V.~Priesemann, ``Inferring change points in the spread of covid-19
  reveals the effectiveness of interventions,'' {\em Science}, 2020.

\bibitem{german-data}
Robert-Koch-Institute,
  ``Rki-covid19,\textit{\url{https://www.arcgis.com/home/item.html?id=f10774f1c63e40168479a1feb6c7ca74}},''
  2020.

\bibitem{R0-calculation}
J.~Heffernan, L.~Wahl, and R.~Smith, ``Perspectives on the basic reproductive
  ratio,'' {\em J R Soc Interface}, vol.~2, p.~281‐293, 2005.

\bibitem{SI-0}
S.~Zhao, P.~Cao, D.~Gao, Z.~Zhuang, Y.~Cai, J.~Ran, M.~K.~C. Chong, K.~Wang,
  Y.~Lou, W.~Wang, L.~Yang, D.~He, and M.~H. Wang, ``{Serial interval in
  determining the estimation of reproduction number of the novel coronavirus
  disease (COVID-19) during the early outbreak},'' {\em Journal of Travel
  Medicine}, 03 2020.

\bibitem{cdc-planning}
NCIRD, ``Covid-19 pandemic planning scenarios,
  \textit{\url{https://www.cdc.gov/coronavirus/2019-ncov/hcp/planning-scenarios.html}},''
  2020.

\bibitem{ICU-bed-data}
{Statistisches-Bundesamt (Destatis)}, ``High hospital bed density in germany
  compared with other countries,
  \textit{\url{https://www.destatis.de/en/press/2020/04/pe20_119_231.html}},''
  2020.

\bibitem{Note2}
Since the the optimal control does not exceed $c = 0.6$ in all our solutions,
  the precise value of this upper bound will not affect our results.

\bibitem{CovidCyclicExit}
O.~Karin, Y.~M. Bar-On, T.~Milo, I.~Katzir, A.~Mayo, Y.~Korem, B.~Dudovich,
  E.~Yashiv, A.~J. Zehavi, N.~Davidovich, R.~Milo, and U.~Alon, ``Adaptive
  cyclic exit strategies from lockdown to suppress covid-19 and allow economic
  activity,'' {\em medRxiv}, 2020.

\bibitem{aron1984seasonality}
J.~L. Aron and I.~B. Schwartz, ``Seasonality and period-doubling bifurcations
  in an epidemic model,'' {\em Journal of theoretical biology}, vol.~110,
  no.~4, pp.~665--679, 1984.

\bibitem{adaptive-epidim}
E.~P. Fenichel, C.~Castillo-Chavez, M.~G. Ceddia, G.~Chowell, P.~A.~G. Parra,
  G.~J. Hickling, G.~Holloway, R.~Horan, B.~Morin, C.~Perrings, {\em et~al.},
  ``Adaptive human behavior in epidemiological models,'' {\em Proceedings of
  the National Academy of Sciences}, vol.~108, no.~15, pp.~6306--6311, 2011.

\bibitem{grenfell-mobility}
C.~O. Buckee, S.~Balsari, J.~Chan, M.~Crosas, F.~Dominici, U.~Gasser, Y.~H.
  Grad, B.~Grenfell, M.~E. Halloran, M.~U. Kraemer, {\em et~al.}, ``Aggregated
  mobility data could help fight covid-19.,'' {\em Science}, vol.~368,
  no.~6487, p.~145, 2020.

\bibitem{grenfell-asymptomatic}
C.~M. Saad-Roy, N.~S. Wingreen, S.~A. Levin, and B.~T. Grenfell, ``Dynamics in
  a simple evolutionary-epidemiological model for the evolution of an initial
  asymptomatic infection stage,'' {\em Proceedings of the National Academy of
  Sciences}, vol.~117, no.~21, pp.~11541--11550, 2020.

\bibitem{hartl1995survey}
R.~F. Hartl, S.~P. Sethi, and R.~G. Vickson, ``A survey of the maximum
  principles for optimal control problems with state constraints,'' {\em SIAM
  review}, vol.~37, no.~2, pp.~181--218, 1995.

\bibitem{US-data}
{The-Covid-Tracking-Project}, ``{\rm USA} historical data,
  \textit{\url{https://covidtracking.com/data/us-daily}},'' 2020.

\bibitem{SI-1}
A.~A. H.~Nishiura, N.M.~Linton, ``{Serial interval of novel coronavirus
  (COVID-19) infections.},'' {\em Int J Infect Dis}, vol.~93, p.~284–286,
  2020.

\bibitem{SI-2}
Y.~W. e.~a. Z.~Du, X.~Xu, ``{Serial interval of COVID-19 from publicly reported
  confirmed cases},'' {\em Emerg Infect Dis}, 2020.

\end{thebibliography}

\end{document}